\documentclass[11pt]{article}
\pdfoutput=1
\usepackage{amsfonts,amsmath,amsthm,graphicx,comment}

\usepackage[margin=1.00in]{geometry}

\usepackage[T1]{fontenc}

\usepackage{wrapfig}
\usepackage{multirow}

\usepackage[normalem]{ulem}
\usepackage{enumitem}

\newcommand\p{\mathrm{P}}
\newcommand\pp{\mathrm{PP}}

\newcommand\np{\mathrm{NP}}
\newcommand\conp{\mathrm{coNP}}\newcommand\pspace{\mathrm{PSPACE}}
\newcommand\npspace{\mathrm{NPSPACE}}

\newcommand\expspace{\mathrm{EXPSPACE}}

\newcommand\nl{\mathrm{NL}}
\newcommand\ac{\mathrm{AC}}
\newcommand\nc{\mathrm{NC}}

\newcommand\sharpp{\#\mathrm{P}}

\newcommand\boolean{\mathbf{N}}
\newcommand\TRUE{\mathit{true}}
\newcommand\FALSE{\mathit{false}}

\newcommand\problem[1]{{\textsc{#1}}}
\newcommand\myemph[1]{\emph{#1}}
\newcommand\config[1]{\vec{#1}}
\newcommand\syst[1]{{#1}}
\newcommand\deftitle[1]{{\bf (#1)~}}

\newtheorem{theorem}{Theorem} 
\newtheorem{corollary}[theorem]{Corollary} 
\newtheorem{proposition}[theorem]{Proposition} 
\newtheorem{definition}{Definition}
\newtheorem{question}{Question}

\newcommand\boxedq[2]{
\smallbreak
{\centering
	\fbox{ \begin{minipage}{0.972\textwidth}
	\begin{question}\label{#1}{#2}\end{question} %
\end{minipage} } }
\smallbreak
}

\usepackage{authblk}

\title{A Theory for Discrete-time Boolean Finite Dynamical Systems with Uncertainty}

\author[1]{Mitsunori Ogihara}
\author[2]{Kei Uchizawa}
\affil[1]{Department of Computer Science, University of Miami, Coral Gables, FL, USA, ogihara@cs.miami.edu}
\affil[2]{Graduate School of Science and Engineering, Yamagata University,  Yonezawa, Yamagata, Japan, uchizawa@yz.yamagata-u.ac.jp}
\date{}

\begin{document}
\sloppy

\hyphenpenalty 10000
\maketitle

\thispagestyle{empty}

\begin{abstract}
\emph{Dynamical Systems} is a field that studies the collective behavior of objects that update their states according to some rules.
Discrete-time Boolean Finite Dynamical System (DT-BFDS) is a subfield where the systems have some finite number of objects whose states are Boolean values, and the state updates occur in discrete time.
In the subfield of DT-BFDS, researchers aim to (i) design models for capturing real-world phenomena and using the models to make predictions and (ii) develop simulation techniques for acquiring insights about the systems' behavior.
Useful for both aims is understanding the system dynamics mathematically before executing the systems.
Obtaining a mathematical understanding of BFDS is quite challenging, even for simple systems, because the state space of a system grows exponentially in the number of objects.
Researchers have used computational complexity to circumvent the challenge.
The complexity theoretic research in DT-BFDS has successfully produced complete characterizations for many dynamical problems.

The DT-BFDS studies have mainly dealt with deterministic models, where the update at each time step is deterministic, so the system dynamics are completely determinable from the initial setting.
However, natural systems have uncertainty.
Models having uncertainty may lead to far-better understandings of nature.
Although a few attempts have explored DT-BFDS with uncertainty, including stochastic initialization and tie-breaking, they have scratched only a tiny surface of models with uncertainty.
The introduction of uncertainty can be through two schemes.
One is the introduction of alternate update functions.
The other is the introduction of alternate update schedules.
This paper establishes a theory of models with uncertainty and proves some fundamental results.
\end{abstract}

\section{Introduction}\label{sec:intro}

Discrete-time Boolean Finite Dynamical System (DT-BFDS)~\cite{kau:k:metabolic-stability} (see, also, e.g.,~\cite{gla-pas:j:stable-oscillations,gla-pas:j:prediction}) is a subfield of dynamical systems. DT-BFDS represents a network of nodes that collectively evolve and is similar to cellular automata~\cite{sut:j:complexity} and Hopfield networks~\cite{hop:j:neural}.

An $n$-node discrete-time Boolean finite dynamical system is a function from the set of $n$-dimensional Boolean vectors to itself.
Starting from an initial vector (or \myemph{initial configuration}) at time $0$, the system at each time step (or \myemph{round}) applies the Boolean function to obtain a new configuration and updates the configuration with it.
There are two major schemes for updating.
The first is \myemph{parallel updating}, where all the nodes update concurrently.
The second is \myemph{sequential updating}, where the nodes update one at a time in some order.
Also, \myemph{asynchronous updating} is a chaotic scheme that allows each node to skip its update.

Researchers use DT-BFDS to study \myemph{network dynamics} in a wide array of disciplines including molecular biology~\cite{alb-rek:j:topology,vel-sti:j:lac,koc-sem:j:collective}, chemistry~\cite{tha-alb:j:immune}, genetics (e.g.,~\cite{pal-sal-ara:j:enumeration,mor-ped:j:stability}, economics~\cite{bha-maj:j:shocks}, and sociology~\cite{chi-lis-etal:c:diffusion}.
By ``network dynamics,'' we mean how a network's state changes over time, starting from an initial configuration.
When the updates are deterministic, a computer simulation is a natural method for examining the evolution of the network.
For example, the authors of~\cite{alb-rek:j:topology} show that a Boolean network can simulate the changes in gene expression occurring in Drosophila \emph{melanogaster}, and the authors of~\cite{vel-sti:j:lac} successfully model the bi-stability of the \emph{lac operon}.

An essential property of DT-BFDS's evolution is that whatever the initial configuration may be, the system will eventually arrive at a configuration on a \myemph{loop} (or \myemph{cycle}).
If the loop is a \myemph{self-loop}, i.e., it has only one configuration, it is a \myemph{fixed point}.
Questions that draw practitioners' interests include, given an initial configuration, how many steps are necessary for the system to arrive at a cycle, at which point the path connects with the cycle, and how long the loop is (the length is one if the cycle is a fixed point).

A DT-BFDS has a natural representation as a directed graph.
Given a DT-BFDS $\syst{F}$, we build its DT-BFDS representation $G = (V, E)$, where $V$ is the set of all configurations and $E = \{ (u, v) \mid u, v \in V \land \syst{F}$ transforms $u$ to $v$ in one time step $\}$.
For a system with $n$ nodes, the size of the graph, $\| V \|$, is $2^n$.
The graph representation enables researchers to study DT-BFDS's properties using graph-theoretic concepts.
Because the configuration graph has an exponential size in the number of nodes, exploring the configuration graph exhaustively can become burdensome quickly.
Researchers began turning to computational complexity to assess the difficulty of finding an answer to graph-theoretic questions about configuration graphs, wth some initial work in~\cite{bar-rei:j:elements-I,bar-mor-rei:j:elements-II,bar-mor-rei:j:elements-III}.
They have found that the complexity of the problems about system dynamics varies widely, depending on the update functions' complexity.

Despite the great progress in the complexity studies of DT-BFDS, ample room appears exist for furthering the study with the use of uncertainty.
The complexity-theoretic investigations have been chiefly on the systems that operate deterministically, and little work currently exists about \myemph{DT-BFDS with uncertainty.}
Considering that uncertainty is always part of the game in the dynamic systems existing in nature, BFDS models with uncertain action choices will naturally extend the existing systems and enrich the field of BFDS.
In fact, recent work~\cite{pau-sen:t:nondet} speaks of the need for more general DT-BFDS, particularly the need for nondeterministic models as unifying tools.
 
There are two places where a system's behavior can become uncertain.
One is that the nodes may not sure of the functions they use.
The other is that the nodes are unsure of the order in which they perform updates.
It appears that different levels of uncertainty exist for both kinds.
Combining the two kinds of uncertainty will give DT-BFDS models with uncertainty.

Given new models with uncertainty, the configuration graphs become more complex.
The nodes of the graphs may now have multiple outgoing edges.
Enumerating all the outgoing edges may be difficult to accomplish in a reasonable amount of time.
Even checking if an arc exists from one configuration to another may be non-trivial.
The new characteristics can bring changes to the computational complexity of some of the graph-theoretic questions.
%
%
Quantification and randomness can be useful for exploring the configuration graph's with multiple outgoing edges.
Using quantification we ask questions like if $a$ has a self-loop (existential), if $a$ has multiple subsequent configurations (universal), how long is the longest simple path from $a$ to $b$ (maximum), and how long is the shortest simple from $a$ to $b$ (minimum).
The use of such quantifiers can bring new characterizations.

Another approach to investigate computing systems with uncertainty is the use of randomness.
Instead of quantifications, we assume some probability distribution on possible actions and ask about the probability that a structural property holds and the expected quantity.
For example, we may ask how likely it is for an update to bring a configuration back to itself and ask for the expected number of steps for a configuration to turn into another.
Stochastic dynamical systems have appeared in real-valued dynamical systems  (e.g.,~\cite{arn:b:random}).
Researchers have used randomness in experimental work, e.g., for perturbing the systems~\cite{zha-kri:j:probabilistic}, activation~\cite{luo-wan:j:dynamics}, generating delays~\cite{cha-our-lim:j:majority-rule}, and tie-breaking~\cite{bha-maj:j:shocks}.
It is a fascinating question what properties of DT-BFDS with uncertainty we can bring out with the user of randomness.

Furthermore, we can study models with two types of uncertainty (function selection and update sequence selection) by combining two quantifiers or combining a quantifier and a probability assessment.
We can study model's robustness through such approaches.

In this paper, we define formally models with uncertainty and study comprehensively the complexity of new models through various types of lenses.

\section{Preliminaries}\label{sec:prelim}

In this section, we give basic definitions of the standard deterministic DT-BFDS studied in the literature. In Section~\ref{sub:dtf-bds}, we explain deterministic DT-BFDSs and their components. In Section~\ref{sub:graphs}, we define dependency and configuration graphs behind DT-BFDSs. In Section~\ref{sub:problems}, we present basic computational problems on deterministic DT-BFDSs, and overview related results. Based on these definitions, we will introduce DT-BFDS with uncertainty in the next section.

\subsection{Deterministic DT-BFDS, updating functions and updating schemes}\label{sub:dtf-bds}

Throughout this paper, $\boolean$ denotes the Boolean basis $\{ 0, 1 \}$, and for all positive integer $m$, $[m]$ denotes the set $\{ 1, \ldots, m \}$, and $\syst{S}_n$ denotes the set of all sequences of $[m]$, that is, the set of all bijections from $[m]$ to itself $[m]$.

Define DT-BFDS as follows.
\begin{definition}\label{def:dtf-bds}
\deftitle{DT-BFDS}
Let $n$ be a positive integer.
An $n$-node DT-BFDS $\syst{F}$ is a tuple of $n$ Boolean functions $(f_1, 
\ldots, f_n)$ that are each from $\boolean^n$ to $\boolean$.
The $n$ functions jointly define a mapping from $\boolean^n$ to itself, where each element of $\boolean^n$ is referred to as a \myemph{configuration}.
\end{definition}
Since investigating nature through simulations is a motivation for DT-BFDS, the update functions for DT-BFDS are simple Boolean functions, such as OR, AND, XOR, XNOR, NOR, and NAND (see, e.g.,~\cite{bar-hun-etal:j:gardens,mad-ros-lal:j:feasibility,ogi-uch:c:synchronous,kaw-ogi-uch:c:generalized}).
The most restricted update functions are \myemph{unary} functions.
\begin{definition}\label{def:unary}
	\deftitle{Unary Functions}
	The \myemph{positive unary} function produces its input without as is.
	The \myemph{negative unary} function produces the negation of its input.
\end{definition}

In the standard DT-BFDS, all nodes must perform one update at each time step.
There are two standard updating schedules:

\begin{definition}\label{def:updating}
	\deftitle{Updating schedule}
	Let $\syst{F} = (f_1, \ldots, f_n)$ be an $n$-node DT-BFDS for some $n \geq 1$.
	Let $\config{c} = (\config{c}_1, \ldots, \config{c}_n)$ be a configuration.
	\begin{enumerate}
		\item
		We say that $\syst{F}$ is a \myemph{parallel} DT-BFDS if its nodes update in parallel (or concurrently).
		In other words, 
		for each configuration $\config{c}$, $\syst{F}$ maps $\config{c}$ to  $(f_1(\config{c}), \ldots, f_n(\config{c}) )$.
		
		\item
		We say that $\syst{F}$ is a \myemph{sequential} DT-BFDS if for some sequence $\pi \in \syst{S}_n$, the update occurs for nodes $\pi(1), \ldots, \pi(n)$ in this order.
		More precisely,  for each $i \in [n]$, let $\tilde{f}_i$ be the \myemph{completion} of $f_i$ with $\boolean^n$ as the domain:
		\[
		\tilde{f}_i(\config{c}_1, \ldots, \config{c}_n) = (\config{c}_1, \ldots, \config{c}_{i-1}, f_i(\config{c}_1, \ldots, \config{c}_n), \config{c}_{i+1}, \ldots, \config{c}_n).
		\]
		Then , for each configuration $\config{c}$, $\syst{F}$ maps $\config{c}$ to
		\[
		\tilde{f}_{\pi(n)}( \cdots ( \tilde{f}_{\pi(2)} ( \tilde{f}_{\pi(1)} ( \config{c} ) ) \cdots ) ).
		\]
	\end{enumerate}
\end{definition}

\paragraph{Encoding}
We assume that an encoding of a DT-BFDS consists of the following:
\begin{itemize}
	\item
	The values of $n$ in unary.
	\item
	The encoding of each function $f_i$ as a Boolean circuit, $i \in [n]$.
	\item
	The updating schedule. In the case of fixed permutation, the permutation the system is to use.
	Each permutation is a sequence of $n$ binary numbers in $[n]$.	
\end{itemize}
We can define a class of DT-BFDS putting restrictions on the following:
\begin{itemize}
	\item
	The structure of the dependency graph, specified with properties like
	\begin{itemize}
		\item
		the maximum/minimum fan-in,
		\item
		the maximum/minimum fan-out,
		\item
		whether the graph is acyclic or not, and
		\item
		whether or not each node must/must not depend on itself (that is, whether a self-loop is permissible in the dependency graph, if so, whether each node must have a self-loop).
	\end{itemize}
	\item
	The types of functions available for the update functions.
	\item
	The types and the number of permutations available for the permutation list.
\end{itemize}

\subsection{Configuration graph, and dependency graph}\label{sub:graphs}

The dependency graph represents the dependency among the nodes.

\begin{definition}\label{def:dependency-graph}
\deftitle{Dependency graph}
Let $V = \{ v_1, \ldots, v_n \}$ be the nodes of a DT-BFDS $\syst{F}$ and let $f_1, \ldots, f_n$ be the update functions of $\syst{F}$.
The \myemph{dependency graph} of $\syst{F}$ is the directed graph $G = (V, E)$ such that
\begin{center}
$E  = \{ (v_i, v_j) \mid$ the function $f_j$ depends on the state value of $v_i \}$.
\end{center}
\end{definition}

The configuration graph of a system specifies the input-output relation.

\begin{definition}\label{def:configuration-graph}
\deftitle{Configuration graph}
Let $\syst{F}$ be an $n$-node DT-BFDS for some $n\geq 1$.
The \myemph{configuration space} of $\syst{F}$ is $\boolean^n$.
The \myemph{configuration graph} of the system $\syst{F}$ is a $2^n$-node directed graph whose directed edges consist of $(\config{c},\config{d})$ such that $\syst{F}(\config{c}) = \config{d}$.
Each node of the configuration graph has an out-degree of $1$.
The configuration graph has a self-loop at each $\config{c}$ wherever $\syst{F}(\config{c}) = \config{c}$.
\end{definition}


\subsection{Properties of configuration graphs}\label{sub:problems}

We can interpret the questions about the network's dynamics as questions about the configuration graph's structure.
Since the configuration graph has $2^n$ nodes, we ask if it is possible to \myemph{answer, without running simulations, if the network has the desired property.}
These questions appear on the next list.

Stating the problems on the list require some definitions.
For an arbitrary configuration $\config{c}$, the path from $\config{c}$ eventually arrives at a node $\config{d}$ so that the path from $\config{d}$ returns to $\config{d}$, i.e., $\config{d}$ is on a cycle.
Amongst all the configurations on the cycle containing $\config{d}$, specific interest is on the point at which the path from $\config{c}$ meets the cycle for the first time.
The meeting point is the \myemph{entry point}.
The path from $\config{c}$ leading to the entry point is the \myemph{tail} from $\config{c}$.
A special kind of cycle is a self-loop, i.e., one with a length of $1$.
A configuration $\config{c}$ with a self-loop satisfies $\syst{F}(\config{c}) = \config{c}$, and such a configuration is called a \myemph{fixed point}.
A configuration $\config{c}$ may have a predecessor, i.e., a configuration $\config{e}$ such that $\syst{F}(\config{e}) = \config{c}$.
A configuration without a predecessor is a \myemph{Garden of Eden}~\cite{ale-dia-etal:j:predecessors,bar-hun-etal:j:gardens}.

Here is a list of structural questions.
On the list, $\config{c}$ and $\config{d}$ are configurations.
The computational problems have their natural decision-problem counterparts with which the computation is possible by way of prefix search or binary search.
\begin{enumerate}
\item
\problem{Reachability}~\cite{bar-hun-etal:j:reachability,nal-rem-et-al:j:consistent}: Is $\config{d}$ reachable from $\config{c}$?
\item
\myemph{$t$-Reachability} for a constant $t\geq 1$: Is $\config{d}$ reachable from $\config{c}$ in at most $t$ time steps?
\item
\problem{Path Length}~\cite{bar-hun-etal:j:reachability}: What is the path length from $\config{c}$ to $\config{d}$? (The length is $-1$ if the path is non-existent.)
\item
\problem{Path Intersection}~\cite{ogi-uch:c:synchronous}: Do the path from $\config{c}$ and the path from $\config{d}$ intersect?
\item
\problem{Tail Length}~\cite{bar-hun-etal:j:reachability,luo-wan:j:dynamics}: What is the tail length of $\config{c}$?
\item
\problem{Garden of Eden}~\cite{ale-dia-etal:j:predecessors,bar-hun-etal:j:predecessor}: Is $\config{c}$ a Garden of Eden?
\item
\problem{$t$-Garden of Eden} for a constant $t\geq 1$: Is there a Garden of Eden from which $\config{c}$ is reachable in exactly $t$ time steps?
\item
\problem{Counting Predecessors}~\cite{ale-dia-etal:j:number-of-solutions}: How many predecessors does $\config{c}$ have? 
\item
\problem{Counting Gardens of Eden}~\cite{hom-kos:j:dichotomy}: How many nodes are Gardens-of-Eden?
\item
\problem{Cycle Point}: Is $\config{c}$ on a cycle?
\item
\problem{Cycle Length}~\cite{bar-hun-etal:j:reachability,luo-wan:j:dynamics,ogi-uch:c:synchronous}: How long is the cycle containing $\config{c}$? (The length is $0$ if $\config{c}$ is not on a cycle.)
\item
\problem{Counting Cycles}~\cite{ara-gom-sal:j:limit}: How many disjoint cycles does the graph have?
\item
\problem{Fixed-point Existence}~\cite{bar-hun-etal:j:gardens,ale-bar-etal:j:fixed-points}: Does the graph have a fixed point?
\item
\problem{Counting Fixed-points}~\cite{hom-kos:j:dichotomy,hei-mal-etal:j:finding-cycles}: How many fixed points are in the graph?
\newcounter{elast1}
\end{enumerate}
Since each edge is polynomial-time computable and the configurations require $n$ bits for representation, we can test the reachability in at most $m$ number of steps in nondeterministic $O(n)$ space.
Then, Savitch's Theorem~\cite{sav:j:relationships} gives that the reachability in deterministic $O(n)$ space.
It follows from this observation that all these problems are solvable in $\pspace$.
Researchers have shown that with proper choice of the update functions, some are $\pspace$-complete, and some are $\np$-complete.
For example,  where both the $2$-fan-in AND and the $3$-fan-in OR are available as update functions, one can design the system so that its configuration graph represents the transitions of configurations of a polynomial space-bounded deterministic Turing machine. The design gives the $\pspace$-hardness of \problem{Reachability}.
Then, as a corollary to the result, one can show that \problem{Path Length}, \problem{Path intersection}, \problem{Tail Length}, \problem{Fixed Point Existence}, and \problem{Cycle Point} are $\pspace$-complete (see, e.g.,~\cite{bar-hun-etal:j:reachability,ale-bar-etal:j:fixed-points,ogi-uch:j:directed}).
Proving the result involves adding modifications to the Turing machine and gadgets.
These results hold as long as the available functions form a Boolean basis.
Also, researchers have shown that \problem{Predecessor Existence} and \problem{$3$-Reachability} are $\np$-complete (e.g.,~\cite{ale-bar-etal:j:fixed-points,ale-dia-etal:j:number-of-solutions}).
The $\pspace$-completeness and the $\np$-completeness results are valid for parallel and sequential models.
Furthermore, \problem{Counting Cycle}, \problem{Fixed Point Counting}, and \problem{Garden of Eden Counting} are $\sharpp$-complete, and their related decision problems are $\pp$-complete~\cite{hom-kos:j:dichotomy,bar-hun-etal:j:predecessor,bar-hun-etal:j:reachability}.

\section{Proposed Models with Uncertainty}\label{sec:models}

In this section, we introduce DT-BFDS with uncertainty. In Section~\ref{sub:multiple-choice}, we define multiple-choice and multiple-updating DT-BFDSs. In Section~\ref{sub:graphs_with_uncertaity}, we define dependency and configuration graphs of DT-BFDS with uncertainty. In Section~\ref{sub:problems_with_uncertainty}, we redefine computational problems for DT-BFDSs. In Section~\ref{sub:organization}, we present the organization of the rest of the paper.

\subsection{Multiple-choice, multiple-updating DT-BFDS, and its taxonomy}\label{sub:multiple-choice}

The first step in constructing a DT-BFDS with uncertainty is to define a system with multiple choices for update functions.

\begin{definition}\label{def:k-choice}
\deftitle{Multiple-choice DT-BFDS}
Let $n$ and $k$ be positive integers.
An \myemph{$n$-node, $k$-choice DT-BFDS}, or an \myemph{$(n, k)$ DT-BFDS}, is a collection $\syst{F} = \{ f_{i,j} \mid i \in [n], j\in [k] \}$, where each element $f_{i,j}$ is a function from $\boolean^n$ to $\boolean$.

Unless we state otherwise, we assume that the positive unary function is always available for all DT-BFDS.
\end{definition}
\noindent
The traditional DT-BFDS is a $1$-choice DT-BFDS. 

When we investigate the computational complexity of structural questions on DT-BFDS, the functions available to the nodes include the identity function unless otherwise stated. A DT-BFDS solely of unary functions sounds too simplistic, but can be appropriate in our uncertainty setting. For example,  we can model relaying points in a signal network where input comes from multiple points and the relaying points pick one of the input signals, e.g., on a first-in, first-out basis.

The definition of multiple-choice DT-BFDS requires a specification of how the system selects functions and what schedule the system uses for updates.

\begin{definition}\label{def:function-choices}
\deftitle{Update function selection}
Let $n$ and $k$ be positive integers.
Let $\syst{F} = \{ f_{i,j} \mid i \in [n], j \in [k] \}$ be an $(n,k)$ DT-BFDS and let $J = [j_1, \ldots, j_n]$ be an $n$-element sequence in $[k]^n$.
Then $\syst{F}$ with $J$ as the function indices, denoted by $\syst{F}[J]$, is the $n$-tuple of functions $(f_{1, j_1}, \ldots, f_{n, j_n})$, which is an $n$-node DT-BFDS.
\end{definition}

Note that the process of selecting update functions is separate from the process of choosing an updating scheme. We will study the following function selection schemes and updating schedules for multiple-choice DT-BFDS.
\begin{itemize}
	\item
	The following types are possible for selecting functions.
	\begin{itemize}
		\item
		\myemph{Fixed Selection~} The system uses the first function for each node and ignores the rest. The system is essentially the same as the standard DT-BFDS.
		\item
		\myemph{Coordinated Selection~} The system selects the index set $J$ from $\{ 1^n, \ldots, k^n \}$ at each time step.
		\item
		\myemph{Individual Selection~} The system selects an index sequence from $[k]^n$ at each time step.
		\item
		\myemph{Semi-coordinated (or Semi-individual) Selection~} The nodes are the disjoint union of some subsets. The selections are in coordination within each subset.
	\end{itemize}
	\item
	The following types are possible for scheduling updates.
	\begin{itemize}
		\item
		\myemph{Parallel or Concurrent Schedule~} All the nodes update concurrently at every time step.
		\item
		\myemph{Fixed-permutation Schedule (Fixed-sequence Schedule)~} The system uses one fixed permutation (or fixed sequence), $\pi$, at every time step.
		\item
		\myemph{Permutation-list Schedule (Sequence-list Schedule)~} The system specifies a list of permutation $\{ \pi_1, \ldots, \pi_m \}$ for some $m\geq 1$. The system chooses one from the $m$ permutations independently at each time step.
		The fixed-permutation schedule is the case where $m=1$.
		\item
		\myemph{Arbitrary-permutation Schedule (Arbitrary-sequence Schedule)~} The system selects one permutation independently at each time step.
		\item
		\myemph{Asynchronous Schedule~} The nodes independently choose whether or not to update at each time step.
	\end{itemize}
\end{itemize}

\paragraph{Encoding}
We can follow the encoding of deterministic DT-BFDS to handle uncertainty, but need extra values:
\begin{itemize}
	\item
	The values of $n$ and $k$ in unary.
	\item
	The encoding of each function $f_{i, j}$ as a Boolean circuit, $i \in [n], j \in [k]$.
	\item
	The function selection scheme.
	In the case of semi-coordinated function selection scheme, the parts among which the function selections occur in coordination.
	In the case of fixed function selection, the fixed selection for each node.
	\item
	The updating schedule.
	In the case of fixed permutation, the permutation the system is to use.
	In the case of permutation list, the list of the permutations.
	Each permutation is a sequence of $n$ binary numbers in $[n]$.
\end{itemize}
Besides restrictions on deterministic DT-BFDS, such as a structure on dependency graphs, we also consider the following restriction: 
\begin{itemize}
	\item The value of $k$, e.g., a constant or a function in $n$ as an upper bound.
\end{itemize}

\subsection{Configuration graph, and dependency graph with uncertainty}\label{sub:graphs_with_uncertaity}
We define dependency and configuration graphs for multiple-choice DT-BFDS, as follows.

\begin{definition}\label{def:nondet-dependency}
\deftitle{The dependency graph of a multiple-choice DT-BFDS}
Let $\syst{F} = \{ f_{i,j} \mid i \in [n], j \in [k] \}$ be an $(n,k)$ DT-BFDS for some $n\geq 1$ and $k\geq 1$.
For each $j\in [k]$, let $G_j = (V, E_j)$ be the dependency graph for the deterministic system $(f_{1,j}, \ldots, f_{n,j})$, where each edge is labeled with the index $j$. Then the \myemph{joint dependency graph} that $\syst{F}$ induces is $G = (V, E_1\cup \cdots \cup E_k)$, where on each edge, its label is the set of all $j$ such that the edge appears in $E_j$.
\end{definition}

\begin{definition}\label{def:nondet-configuration}
\deftitle{The configuration graph of a multiple-choice DT-BFDS}
Let $n$ and $k$ be positive integers.
Let $\syst{F} = \{ f_{i,j} \mid i \in [n], j \in [k] \}$ be an $(n, k)$ DT-BFDS with a specific updating schedule and a specific function selection scheme.
Define the configuration graph of $\syst{F}$ as the graph $G= (V, E)$, where there is an arc from a configuration $\config{c}$ to a configuration $\config{d}$ if $\config{d}$ is one of the possibilities $\syst{F}$ generates in one step, given $\config{d}$ as the input.
In addition, each directed edge of the configuration graph has a label, which is the set of all choices in the permissible actions that achieve the corresponding transition.
Such choices may include update functions and updating sequences.
\end{definition}


\subsection{Properties of configuration graphs with uncertainty}\label{sub:problems_with_uncertainty}

With the new definition of configuration graphs, we need to redefine the structural questions so that the existence of an arrow is an $\np$ question.
\begin{enumerate}
\item
\problem{Reachability}: Is there a path from $\config{c}$ to $\config{d}$?
\item
\problem{$t$-Reachability}: Is there a path from $\config{c}$ to $\config{d}$ whose length is at most $t$?
\item
\problem{Minimum/Maximum Path Length}: How long is the shortest/longest simple path from $\config{c}$ to $\config{d}$?
\item
\problem{Path Intersection}: Do \emph{any} path from $\config{c}$ and \emph{any} path from $\config{d}$. intersect?
\item
\problem{Tail Length}: How long is the shortest path from $\config{c}$ to \emph{any} cycle?
\item
\problem{Garden of Eden}: Is $\config{c}$ a Garden of Eden?
\item
\problem{$t$-Garden of Eden}: Is a Garden of Eden reachable from $\config{c}$ in $t$ backward steps?
\item
\problem{Counting Predecessors}: How many configurations are the predecessors of $\config{c}$?
\item
\problem{Counting Gardens of Eden}: How many nodes are Gardens of Eden?
\item
\problem{Cycle Point}: Is $\config{c}$ on \emph{any} cycle?
\item
\problem{Minimum/Maximum Cycle Length}: How long is the shortest/longest simple cycle that goes through $\config{c}$?
\item
\problem{Counting Cycles}: How many simple cycles go though $\config{c}$?
\item
\problem{Fixed-Point Existence}: Does the graph have a node with a self-loop?
\item
\problem{Counting Fixed Points}: How many nodes have self-loops?
\item
\problem{Counting Subsequent Configurations}: How many different subsequent configurations does $\config{c}$ have?
\item
\problem{Complete Fixed-Point}: Is $\config{c}$ a complete fixed-point?
\item
\problem{Complete Fixed-Point Existence}: Does the graph include a complete fixed-point?
\item
\problem{Counting Complete Fixed-Points}: How many complete fixed-points are there?
\item
\problem{Counting Paths}: How many simple paths exist from $\config{c}$ to $\config{d}$?
\end{enumerate}

\subsection{Organization of the rest of the paper}\label{sub:organization}
The rest of the paper is organized a follows. In Section \ref{sec:general}, we provide general upperbounds that hold regardless of the updating schedule. In Section~\ref{sec:cross-model}, we ask if a particular combination of a selecting function and a scheduling update includes another one by means of simulation, and provide several inclusion results. In Section~\ref{sec:parallel-independent}, we consider DT-BFDSs with parallel or fixed-permutation updating combined with individual function selection, and show upper and lower bounds on $t$-\problem{Reachability} and its related problems. In Section~\ref{sec:coordinated}, we consider DT-BFDSs with parallel updates with the coordinated function selections, and investigate a relationship between \problem{Reachability} and the graph isomorphism problem. In Section~\ref{sec:nondeterministic-sequential-updating}, we consider DT-BFDSs with permutation-list and arbitrary-permutation updating, and show complexity results on $t$-\problem{Reachability} and its variants. In Section~\ref{sec:open}, we conclude with some open problems.

The following table summarizes the sections where the combinations of updating schedules and selection schemes are considered.
\begin{table}[bth]
		\begin{center}
			\begin{tabular}{|c|c|c|c|c|}
				\hline
				\multirow{2}{*}{updating schedule} & \multicolumn{4}{|c|}{function selection} \\
				\cline{2-5}
				& fixed & coordinated & individual & semi-coordinated \\
				\hline
				asynchronous & - & - & - & - 
				\\
				\hline
				parallel & - & 7 & 6  & -
				\\
				\hline
				fixed permutation & - & - & 6  & -
				\\
				\hline
				permutation list & - & - & 8 & -
				\\
				\hline
				arbitrary permutation & - & - & 8 & -
				\\
				\hline
			\end{tabular}
		\end{center}
		\caption{The numbers indicate which section the mode is considered. Since Sections~\ref{sec:general} and \ref{sec:cross-model} are related to all the types of DT-BFDS, we omit them.}
		\label{t:work}
\end{table}


\section{General upper bounds}\label{sec:general}

The problem \problem{$t$-Reachability} in the standard DT-BFDS model is in $\p$ when $t$ is bounded by  some polynomial in $n$. In the case of DT-BFDS with uncertainty, configurations may have multiple possible subsequent configuration.
The number of possibilities can be exponential in the number of nodes if either the system uses an arbitrary sequence for scheduling updates or selects the function individually for the nodes.
In such a case, the problem of testing an edge's existence is in $\np$.

\begin{proposition}\label{prop:edge-existence}
\problem{$1$-Reachability} is in $\np$.
\end{proposition}

\begin{proof}
Let $\syst{F}$ be an arbitrary $(n, k)$ system for some $n, k \geq 1$.
Let $\config{c}$ and $\config{d}$ be two configurations of the system.
We observe the following.
\begin{itemize}
\item
In the case when the update schedule is permutation list having length $L$, we have only to examine each permutation of the list, which gives a multiplicative factor of $L$ to the running time.
\item
In the case when the update schedule is arbitrary permutation, we can in $O(n^2)$ time select a permutation nondeterministically.
\item
In the case when the update schedule is asynchronous, we need to select the nodes that perform an update in the round, separate them into groups, and then select an order among the groups.
We can accomplish the selection as follows:
\begin{enumerate}
\item
For each $i \in [n]$, nondeterministically select whether or not to perform an update on the $i$-node.
\item
Nondeterministically permute the sequence of indices $i$ we have chosen in the first step.
\item
Let $[ i_1, \ldots, i_r ]$ the permutation we have obtained in the previous step and let $[u_1, \ldots, u_r ]$ be the nodes that these indices indicate.
Start the first group only with $u_1$ is as its member.
For each $j, 2\leq j\leq r$, nondeterministically select whether or not to place $v_j$ in the same group as $v_{j-1}$.
If the answer is negative, start a new group with $u_j$ as its member.
\end{enumerate}
\item
In the case when the function selection scheme is coordinated, we have only to make an exhaustive search over the $k$ possibilities.
\item
In the case when the function selection is individual, we have only select an update index series in $O(n)$ time.
\end{itemize}
Using the selections made, we check whether the functions and the schedule produce $\config{d}$ from $\config{c}$.
\end{proof}

In light of the observation, we obtain the following computational upper bound for each problem from Section~\ref{sub:problems_with_uncertainty}.

\begin{proposition}\label{prop:general-upper-bound}
The following upper bounds hold regardless of the updating schedule.
In the case where the update schedule is one of parallel, fixed-permutation, and permutation-list and the function selection is either fixed or coordinated, the complex, the upper bounds become those appearing in parentheses.
\begin{enumerate}
\item
$\np$ ($\p$): \problem{$t$-Reachability} for a fixed $t$,  and \problem{Fixed-Point Existence}.
\item
$\conp$: \problem{Garden of Eden} and \problem{Complete Fixed-Point}.
\item
$\Sigma^p_2$ ($\np$): \problem{$t$-Garden of Eden} for a fixed $t$ and \problem{Complete Fixed-Point Existence}.
\item
$\sharpp^{\np}$ ($\sharpp$): \problem{Counting Fixed Points} ($\sharpp$), \problem{Counting Subsequent Configurations}. \problem{Counting Gardens of Eden}, \problem{Counting Predecessors} ($\sharpp$), and \problem{Counting Complete Fixed-Points}.
\item
$\pspace$: \problem{Reachability}, \problem{Minimum/Maximum Path Length}, \problem{Path Intersection}, \problem{Tail Length}, \problem{Cycle Point}, and \problem{Minimum/Maximum Cycle Length}.
\item
$\expspace$: \problem{Counting Cycles} and \problem{Counting Paths}.
\end{enumerate}
\end{proposition}

\begin{proof}
Using the fact that \problem{$1$-Reachability} is in $\np$, we prove the upper bounds as follows.
\begin{enumerate}
\item
For \problem{$t$-Reachability}, we have only to select $t-1$ intermediate configurations nondeterministically and then verify using the verification method for \problem{$1$-Reachability}, if each edge is valid.
For \problem{Fixed-Point Existence}, we have only to select a configuration nondeterministically and verify if the configuration is reachable from itself using the verification method for \problem{$1$-Reachability}.
Thus, these problems are in $\np$.
\item
For \problem{Garden of Eden}, note that the predecessor existence can use a guess of a predecessor and the verification.
A Garden of Eden is a node for which the predecessor-existence test fails, and so the problem is in $\conp$.
For \problem{Complete Fixed-Point}, we can check if all the valid update schedules and the function selections take back to the original.
Thus, the problem is in $\conp$.
\item
For \problem{$t$-Garden of Eden}, we can guess a point to be a $t$-Garden of Eden and then verify that there is a length-$t$ to the input configuration and the point is a Garden of Eden.
For \problem{Complete Fixed-Point Existence}, we have only to select a candidate configuration nondeterministically and then verify that the candidate is a complete fixed-point.
This puts the two problems in $\Sigma^p_2$.
\item
Based on the previous two observations, we have that the counting problems \problem{Counting Fixed Points} ($\sharpp$), \problem{Counting Subsequent Configurations}. \problem{Counting Gardens of Eden}, \problem{Counting Predecessors} ($\sharpp$), and \problem{Counting Complete Fixed-Points} can use a nondeterministic Turing machine that guesses a configuration uniquely and then verifies the requisite property using an $\np$-oracle.
Thus, they are in $\sharpp^{\np}$.
\item
We can test the reachability by dynamically exploring a path to the target by guessing the next point on the path one after another.
For each next point we generate, we check the guess is reachable from the present point in $1$ time step using the aforementioned verification method.
Since there are $2^n$ configurations, we can limit the number of steps to $2^n$.
This means that the reachability is in $\npspace^{\np} = \npspace$.
Since $\npspace = \pspace$ due to Savitch's Theorem~\cite{sav:j:relationships}, we have that the problem is in $\pspace$.
Using the result, we can check reachability in any number of time steps $\leq 2^n$ in $\pspace$, and so the maximum/minimum reachability problems are in $\pspace$.
The cycle point and cycle length problems use the same idea.
To compute the tail length, we have only to check for each configuration, if it is a cycle point and compute the minimum path length to the point, then obtain the minimum of all the path lengths.
\item
For counting simple cycles, we can use the following strategy:
for $\ell = 1, \ldots, 2^n$, we generate all sequences of $\ell$ configurations,
$v_1, \ldots, v_{\ell}$. For each sequence, we check if the last node and the first are identical to each other, there are no other duplication, and if $v_{j}$ is reachable form $v_{j-1}$ in $1$ time step for all $j, 2\leq j \leq \ell$. If the test passes, the sequence is a cycle.
We count the length-$\ell$ sequences passing the test and divide the count by $\ell$.
The result of the division is the number of simple cycles having length $\ell$.
We have only to sum all these cycle counts to obtain the total count of simple cycles.
Writing down a sequence requires $O(2^n)$ space.
Thus, the problem is in $\expspace$.
\end{enumerate}
This proves the proposition.
\end{proof}

A general question arising immediately from the above results is if a model includes another.

\section{Cross-model simulations}\label{sec:cross-model}

Given the various types of DT-BFDS with uncertainty, we question if any new model includes another.
The following holds from the definition.

\begin{definition}\label{def:embedding}
\deftitle{Embedding}
An \myemph{embedding} of a configuration graph $G = (V,E)$ to another configuration graph $G' = (V',E')$ is a one-to-one mapping $\nu: V \rightarrow V'$ with the following property holds for all $a, b \in V$: there is a nontrivial path (i.e., having length $>0$) from $a$ to $b$ in $G$ if and only if there is a nontrivial path from $\nu(a)$ to $\nu(b)$ in $G'$.
The edge expansion rate of the embedding is
\[
\max_{a, b \in V} \left\{ \frac{ \mathrm{the~length~of~the~shortest~nontrivial~path~from~}\nu(a)\mathrm{~to~}\nu(b)}{ \mathrm{the~length~of~the~shortest~nontrivial~path~from~}a\mathrm{~to~}b } \right\}.
\]
\end{definition}

We define simulations using the concept of embedding.

\begin{definition}\label{def:simulation}
\deftitle{Simulation}
Let $\rho > 0$ be a real number.
Let $\syst{S}$ and $\syst{T}$ be two DT-BFDS's.
We say that $\syst{T}$ simulates $\syst{S}$ with an expansion rate $\rho$ if there is an embedding of the configuration graph of $\syst{S}$ into the configuration graph of $\syst{T}$ with its edge expansion rate no more than $\rho$. 
\end{definition}

\begin{definition}\label{def:reduction}
\deftitle{Reduction}
Let $\rho$ be a function from a set of the positive integers to itself.
Let $\mathcal{C}$ and $\mathcal{D}$ be classes of DT-BFDS.
We say that $\mathcal{C}$ is polynomial-time reducible to $\mathcal{D}$ with expansion rate of $\rho(n)$ if there is a pair of polynomial-time functions $(g, h)$ such that for all $n\geq 1$ and for all BT-BFDS's $\syst{S}$ in $\mathcal{C}$ having $n$ nodes, the following properties hold:
\begin{itemize}
\item
The value of $g(\syst{S})$ is a system in $\mathcal{D}$.
\item
For all configurations $\config{c}$ of $\syst{S}$, the value of $h(\syst{S}, \config{c})$ is a configuration of $g(\syst{S})$.
\item
The function $h(\syst{S}, \cdot)$ serves as an embedding of $\syst{S}$ into $g(\syst{S})$ with its edge expansion rate no more than $\rho(n)$.
\end{itemize}
\end{definition}

\begin{proposition}\label{prop:trivial}
For each updating schedule, we have that:
\begin{center}
fixed selection $\subseteq$ coordinated selection $\subseteq$ semi-coordinated selection\\
fixed selection $\subseteq$ individual selection $\subseteq$ semi-coordinated selection
\end{center}
where $\subseteq$ means ``is part of.''
\end{proposition}

What can we say about the updating schemes?
We know very little right now.
Trivially, the fixed-permutation scheme is a special case of the permutation-list scheme.
We can also say that the asynchronous scheme is part of the parallel scheme as follows:

\begin{theorem}\label{thm:asynchronous}
Let $n$ and $k$ be positive integers.
\begin{enumerate}
\item
Each $(n, k)$ DT-BFDS that updates asynchronously and uses individual function selections is simulate-able by an $(n, k+1)$ DT-BFDS that updates in parallel and uses individual function selections with the edge expansion rate equal to $n$.
\item
Each $(n, k+1)$ DT-BFDS that updates in parallel and uses individual function selections is simulate-able by an $(n, k)$ DT-BFDS that updates asynchronously and uses individual function selections with the edge expansion rate equal to $1$, if the identify function is among the function choices for each node.
\end{enumerate}
\end{theorem}

\begin{proof}
Let $n$ and $k$ be positive integers.
\begin{enumerate}
\item
Let $\syst{F} = \{ f_{i j} \mid i \in [n], j \in [k] \}$ be an $(n, k)$ DT-BFDS that updates asynchronously and uses individual function selections.
Let $V = \{ v_1, \ldots, v_n \}$ be the nodes of $\syst{G}$ (and thus, of $\syst{F}$).
For each $i \in [n]$, let $f_{i, k+1}$ be the identity function for $v_i$.
Define $\syst{G} = \{ f_{ij } \mid i \in [n], j \in [k+1] \}$ be an $(n, k)$ DT-BFDS that updates in parallel.
Let $[S_1, \ldots, S_m]$ be an arbitrary sequence of mutually-disjoint nonempty subsets of $[n]$. Because of the mutual disjointness, $m\leq n$.
Suppose, in a time step, updates occur on the nodes with indices in $S_1, \ldots, S_m$ in this order such that the updates occur concurrently among the nodes with their index in $S_h$, but not with 
Let $T = S_1 \cup \cdots \cup S_m$ and $T' = [n] - T$.
For each $i \in T$, let $e_i$ be the index of the function the system has chosen for $v_i$ in this round.
For each $h \in [m]$, define the function selection index sequence $J_h = [ j^{(h)}_1, \ldots, j^{(h)}_n ]$ to be
\begin{center}
$j^{(h)}_i = e_i$ if $i \in S_h$ and $k+1$ otherwise.
\end{center}
Then, for each $h \in [m]$, $\syst{G}[J_j]$ acts as the action $\syst{F}$ for $S_h$ in this round.
Define the embedding $\nu$ of the configurations to be the identity function.
With the embedding $\nu$, for each edge $(a, b)$ in the configuration graph of $\syst{F}$, there is a path having length at most $m$ from $\nu(a)$ to $\nu(b)$ in $\syst{G}$.
This means that if there is a path from $a$ to $b$ having length $\ell$ in the configuration graph of $\syst{F}$, there is a path from $\nu(a)$ to $\nu(b)$ having length at most $\ell m \leq \ell n$ (since $m\leq n$) in the configuration graph of $\syst{G}$.
Thus, the edge expansion rate is at most $n$.

On the other hand, suppose there is an edge $(u, v)$ in $\syst{G}$'s configuration graph.
Let $J = [ j_1, \ldots, j_n ]$ be an index selection sequence that achieves this transition.
Let $S_0 = \{ i \mid j_i \neq k + 1 \}$ and let $J'$ be the sequence we obtain from $J$ keeping only those elements at the positions in $S_0$.
Consider a time step in which the system $\syst{F}$ chooses to update only nodes $v_i$ such that $i \in S_0$ and to update those nodes concurrently.
Then, $\syst{F}$ transitions from $u = \nu^{-1}(u)$ to $v = \nu^{-1}(v)$ in one time step.
This means that every edge of $\syst{G}$ is realization of a possible action in one time step in $\syst{F}$.

This proves the claim.
\item
Let $\syst{G} = \{ f_{i, j} \mid i \in [n], j \in [k+1] \}$ be an $(n, k+1)$ DT-BFDS that updates in parallel with with individual functions selections such that one of the functions for each node is an identity function.
Since $\syst{G}$ makes individual function selections, the order which the functions for any node is permutable without changing the system's behavior.
We thus may assume that $f_{i,k+1}$ is the identity function for all $i \in [n]$.
Define $\syst{F} = \{ f_{i, j} \mid i \in [n], j \in [k] \}$ to be the $(n, k)$ DT-BFDS that updates asynchronously and selects functions individually.
We then follow the proof from the previous part to show that there is a path from $a$ to $b$ in $\syst{G}$'s configuration graph if and only if there is a path from $a$ to $b$ in $\syst{F}$'s configuration.
The edge expansion rate is $1$ for this case, however, since edge on $\syst{G}$ reflects one group's concurrent action and so $\syst{F}$ can simulate in one time step.
\end{enumerate}
This proves the theorem.
\end{proof}

We obtain the following corollary to the theorem.

\begin{corollary}\label{coro:asynch-reduction}
For all integers $k\geq 1$, the class of all $k$-choice DT-BFDS that updates asynchronously and uses individual function selections is polynomial-time reducible to the class of all $(k+1)$-choice DT-BFDS that updates in parallel and uses individual function selections.
\end{corollary}

A significant difference conspicuously exists between the parallel and fixed-permutation updates.
With parallel updating, two nodes, say $u$ and $v$, can exchange their values in one time step, with $u$ having $0$ receiving $1$ from $v$ and simultaneously giving its $0$ to $v$.
Such a one-step state exchange is impossible for sequential updates because one update must occur before the other.
Is the sequential updating less powerful than the parallel updating?
The answer appears to be negative.

\begin{theorem}\label{thm:parallel-to-sequential-reduction}
Suppose that the positive unary function is available for using as update functions.
For all positive integers $n$ and $k$, each $(n, k)$ DT-BFDS that uses parallel updates and makes individual selections is simulate-able by a $(2n, k)$ DT-BFDS that uses sequential updates that uses the same function selection scheme.
The edge expansion rate of the simulation is $1$.
\end{theorem}

\begin{proof}
Let $\mathcal{F} = \{ f_{i, j} \mid i \in [n], j\in [k] \}$ be an $(n, k)$ DT-BFDS that updates in parallel and uses a function selection scheme $\tau$.
Let $v_1, \ldots, v_n$ be the nodes of $\mathcal{F}$ and, by abuse of notation, their state values.
We construct a new $2n$-node DT-BFDS $\mathcal{G} = \{ g_{i, j}  \mid i \in [2 n], j\in [k] \}$ as follows.
\begin{itemize}
\item
We introduce $n$ nodes $v_{n+1}, \ldots, v_{2n}$.
\item
For each $i \in [n]$ and for each $j \in [k]$, $g_{i, j}  = f_{i-n, j}$.
In other words, for all $\ell \in [n]$ and $j \in [k]$, we use $f_{\ell, j}$ to determine the value of $f_{\ell + n}$.
\item
For each $i \in [n]$ and for each $j \in [n]$, $g_{i, j}$ is the positive unary function that takes the value from $v_{i + n}$.
\item
The permutation prescribing the update order is $[n+1, \ldots, 2n, 1, \ldots, n ]$.
\end{itemize}
It is not difficult to see that in each time step, the result of executing one time step of $\mathcal{F}$ appears in the nodes $v_{n+1}, \ldots, v_{2n}$, and then the system collectively copies the values of $v_{n+1}, \ldots, v_{2n}$ to $v_1, \ldots, v_{n}$.

Thus, after each time step, the configuration of $\syst{G}$ is of the form $\config{c} \cdot \config{c}$ for some configuration $\config{c}$ of $\syst{F}$, where $\cdot$ means the vector concatenation.
Define $\nu$ to be the embedding that maps $u$ to $u\cdot u$.
Then, for each configuration pairs $(a, b)$ of $\syst{F}$, there is a path from $a$ to $b$ having length $\ell$ in $\syst{F}$ if and only if there is a path from $\nu(a)$ to $\nu(b)$ in $\syst{G}$ having length $\ell$, where $\ell$ is an arbitrary positive integer.
This means that the edge expansion is $1$.
This proves the theorem.
\end{proof}

\begin{theorem}\label{thm:sequential-to-parallel-reduction}
Suppose that the positive unary function is available for using as update functions.
For all positive integers $n$ and $k$, each $(n, k)$ DT-BFDS that uses sequential updates and makes coordinated selections is simulate-able by a $(k(n+1)^2+1, 2k)$ DT-BFDS that uses parallel updates and makes coordinated selections.
The edge expansion rate of the simulation is $n+1$.
\end{theorem}

\begin{proof}
Let $\syst{S} = \{ f_{i, j} \mid i \in [n], j \in [k] \}$ be an $(n, k)$ DT-BFDS that uses a sequence $\pi$ for update scheduling and makes coordinated updates.
Let $m = k(n+1)^2 + 1$.

We define a system $\syst{T}$ that makes parallel, coordinate updates as follows.
\begin{itemize}
\item
The nodes of $\syst{T}$ are $v_{\ell, i}, \ell \in [n+1], i \in [n]$, $a_{j, \ell}, j \in [k], \ell \in [n+1]$, and a single node $z$.
\item
Each configuration $\config{c} = (c_1, \ldots, c_n)$ of $\syst{S}$ corresponds to  the configuration $\nu(\config{c})$ of $\syst{T}$.
The states of $\nu(\config{c})$ is as follows:
\begin{itemize}
\item
For all $\ell \in [n+1]$, the values of $v_{\ell, i}, i \in [n]$, are identical to $\config{c}$.
\item
For all $j \in [k]$ and $\ell \in [n+1]$, $a_{j, \ell} =1$ if $\ell = 1$ and $0$ otherwise.
\item
The value of $z$ is $0$.
\end{itemize}
\item
There are two $k$-element groups of updating functions $G_j$ and $H_j$, $j\in [k]$.
The total number of updating functions is thus $2k$.
The action of $G_j$, $j \in [k]$, is as follows.
\begin{itemize}
\item
For all $\ell \in [n]$ and $i \in [n]$, $G_j$ copies the state from $v_{\ell, i}$ to $v_{\ell+1, i}$, except that, instead of copying, $G_j$ sets the value of $v_{i\ell+1, \pi(i)}$ to the result of computing $f_{\pi(i), j}$ with the inputs from the $n$ state values $v_{\ell, 1}, \ldots, v_{\ell, n}$.
\item
For all $i \in [n]$, $G_j$ preserves the states of $v_{1, 1}, \ldots, v_{1, n}$ .
\item
For all $\ell \in [n]$, $G_j$ copies the state from $a_{j, \ell}$ to $a_{j, \ell+1}$ and copies the state $0$ from $z$ to $a_{j,1}$.
\item
For all $j' \in [k] - \{ j \}$ and $\ell \in [n+1]$, $G_j$ preserves the state of $a_{j',\ell}$.
\item
$G_j$ preserves the state of $z$.
\end{itemize}
The action of $H_j$, $j \in [k]$, is as follows.
\begin{itemize}
\item
For all $\ell \in [n]$ and $i \in [n]$, $H_j$ copies the state from $v_{n+1, i}$ to $v_{\ell, i}$.
\item
For all $i \in [n]$, $H_j$ preserves the state of $v_{\ell+1, i}$.
\item
For all $\ell, 2\leq \ell \leq n + 1$, $H_j$ copies the state $0$ from $z$ to $v_{j, \ell}$.
It also copies the state from $v_{j, n+1}$ to $v_{j, 1}$.
\item
For all $j' \in [k] - \{ j \}$ and $\ell, 2\leq \ell \leq n + 1$, $H_j$ copies the state $0$ from $z$ to $v_{j', \ell}$.
\end{itemize}
\end{itemize}
Let us examine the actions of $H_1, \ldots, H_k$.
The required pattern in the $a$'s is all $0$ but $a_{1,1}, \ldots, a_{k,1}$; that is,
\begin{center}
\begin{tabular}{cccc|c}
$a_{j,1}$ & $a_{j,2}$ & $\cdots$ & $a_{j,n+1}$ & $j$ \\
\hline  
1 & 0 & $\cdots$ & 0 & $1$ \\
1 & 0 & $\cdots$ & 0 & $2$ \\
$\vdots$ & $\vdots$ & $\vdots$ & $\vdots$ & $\vdots$ \\
1 & 0 & $\cdots$ & 0 & $k$ \\ 
\end{tabular}
\end{center}
When the system applies $H_{j_0}$ for some $j_0 \in [k]$, for the system to change the $a$'s back to the required form, the pattern must be
\begin{center}
\begin{tabular}{cccc|c}
$a_{j,1}$ & $a_{j,2}$ & $\cdots$ & $a_{j,n+1}$ & $j$ \\
\hline  
1 & - & $\cdots$ & - & $1$ \\
$\vdots$ & $\vdots$ & $\vdots$ & $\vdots$ & $\vdots$ \\
1 & - & $\cdots$ & - & $j_0 - 1$ \\
- & - & $\cdots$ & 1 & $j_0$ \\
1 & x & $\cdots$ & - & $j_0 + 1$ \\
$\vdots$ & $\vdots$ & $\vdots$ & $\vdots$ & $\vdots$ \\
1 & - & $\cdots$ & - & $k$ \\ 
\end{tabular}
\end{center}
Here - means ``arbitrary.''
An application of a $G_j$ for any $j$, does not increase the number of $1$'s in any row.
Specifically, for all rows other than the $j$-th one, $G_j$ preserves it, and for the $j$-th row, $G_j$ shifts its $1$ to the right and inserts a $0$ the column position $1$.
If the $1$ is at the column position $n+1$, the $1$ disappears and the $j$-th row becomes all $0$.

From these we observe the following:
\begin{itemize}
\item
Achieving the required format for the $a$'s is only possible by applying some $G_j$ consecutively exactly $n$ times and then applying $H_j$.
\item
With $n$ applications of $G_j$, the $(n+1)$-st row of $v$ becomes the result of applying $f_{i, j}, i \in [n]$ to the first row of $v$.
\item
Following this, an application of $H_j$, copies the $(n+1)$-st row of $v$ to all other rows of $v$, making the system ready for applying another series, consisting of $n$ $G_{j'}$ and $H_{j'}$ for some $j'$.
\end{itemize}
Thus, in $n+1$ time steps, $\syst{G}$ can simulate one step of $\syst{F}$, and that is the only way it can simulate the action of $\syst{F}$ under the constraint that the $a$-part needs preservation.
The edge-expansion rate is $n+1$.
This proves the theorem.
\end{proof}

\begin{corollary}\label{coro:permutation-list-to-parallel}
Suppose that the positive unary function is available for using as update functions.
For all positive integers $L$, $n$, and $k$, each $(n, k)$ DT-BFDS that uses a permutation list consisting of $L$ permutations for sequential updates and makes coordinated selections is simulate-able by a $((k L + n)(n+1) + 1,  2 k L)$ DT-BFDS that uses parallel updates and makes coordinated selections.
The edge expansion rate of the simulation is $n+1$.
\end{corollary}

\begin{proof}
We have only to create $L$ copies of the $a$ part from the above proof, dedicated to the $L$ permutations while sharing the $v$'s and $z$ among the copies.
\end{proof}

The above two results thus give:
\begin{corollary}\label{coro:subsumption} {~}
\begin{itemize}
\item
For coordinated selection scheme,
fixed-permutation $\subseteq$ permutation-list $\subseteq$ parallel.
\item
For individual selection scheme, parallel $\subseteq$ fixed-permutation $\subseteq$ permutation-list.
\item
For the individual function selection scheme, 
asynchronous $\subseteq$ parallel and
parallel $\subseteq$ asynchronous.
\end{itemize}
\end{corollary}

Another interesting question is about the number of choices $k$.
Specifically, is $k$ a parameter that governs the computational complexity of the structural problems?
A ``normal'' form of nondeterministic Turing machine has at most two possible moves. It is possible to convert an arbitrary nondeterministic Turing machine to its ``normal'' form. The conversion is at the cost of constant slow down.

\begin{theorem}\label{thm:coordinated-k-to-3}
Let $n \geq 1$ and $k \geq 3$.
For each $(n, k)$ system $\syst{F} = \{ f_{i, j} \mid i \in [n], j \in [k] \}$ that uses an update scheme $U$ and updates in coordination, there exists an $(kn + 3, 3)$ system $\syst{G} = \{ g_{i, j} \mid i \in [k n + 3], j \in [3] \}$ that simulates $\syst{F}$ that uses the same update scheme as $U$ an updates in coordination.
The edge expansion rate of the simulation is $k+1$.
\end{theorem}

\begin{proof}
Let $n \geq 1$ and $k \geq 3$.
Let $\syst{F} = \{ f_{i, j} \mid i \in [n], j \in [k] \}$ be an $(n, k)$ system that uses an update scheme $U$ and updates in coordination.
Let $v_1, \ldots, v_n$ be the nodes of $\syst{F}$.
Let $\config{c} = (c_1, \ldots, c_n)$ and $\config{d} = (d_1, \ldots, d_n)$, which are the initial configuration and the target configuration, respectively.
Suppose that the scheme $U$ is parallel.
We define a system $\syst{G} = \{ g_{i, j} \mid i \in [k n + 3], j \in [3] \}$ as follows:
\begin{itemize}
\item
The nodes of $\syst{G}$ is $v_{i, j}, i \in [i], j \in [k]$, and $y_1, y_2, y_3$.
The node order is
\[
v_{1, 1}, \ldots, v_{n, 1}, v_{1, 2}, \ldots, v_{n, 2}, \ldots, v_{1, k}, \ldots, v_{n, k}, y_1, y_2, y_3
\]
In other words, it is $k$ copies of  $v_1, \ldots, v_n$ followed by $y_1, y_2, y_3$.
We call auxiliary nodes.
\item
The initial configuration, $\config{s}$, is $k$ copies of $\config{c}$ followed by $0, 1, 1$.
\item
The target configuration, $\config{t}$, is $\config{d}$ followed by $(k-1)$ copies of $0^n$ then by $0, 0, 1$.

\item
The first group of functions $g_{\ell, 1}, \ell \in [k n + 3]$, works as follows:
\begin{itemize}
\item
For each $j \in [k]$, we update $j$-th copy with the $j$-th function group $f_{1, j}, \ldots, f_{n, j}$.
\item
We apply the identity function to $y_1$ and $y_2$ and copy $y_2$'s state to $y_3$.
\end{itemize}

\item
The second group of functions $g_{\ell, 2}, \ell \in [k n + 3]$, works as follows:
\begin{itemize}
\item
For each $j \in [k]$, we copy from Group $j$ to Group $j+1$, where we use $1$ in the case of $j + 1= k + 1$.
\item
We apply the identity function to $y_1$ and $y_2$ and copy $y_2$'s state to $y_3$.
\end{itemize}

\item
The third group of functions $g_{\ell, 3}, \ell \in [k n + 3]$, works as follows:
\begin{itemize}
\item
We preserve the states of the first group.
\item
We copy the state from $y_1$ to each node of the other $(k - 1)$ groups.
\item
We preserve $y_1$'s state, copy $y_1$'s state to $y_2$, and copy $y_2$'s state to $y_3$.
\end{itemize}
\end{itemize}
Note that the three function groups have the following effect on the auxiliary nodes.

\begin{center}
\begin{tabular}{|c|c|c|c|}
\hline
& \multicolumn{3}{|c|}{input} \\
\cline{2-4}
& $(0, 1, 1)$ & $(0, 0, 1)$ & $(0, 0, 0)$ \\
\hline
Function group 1 & $(0, 1, 1)$ & $(0, 0, 0)$ & $(0, 0, 0)$ \\
\hline
Function group 2 & $(0, 1, 1)$ & $(0, 0, 0)$ & $(0, 0, 0)$ \\
\hline
Function group 3 & $(0, 0, 1)$ & $(0, 0, 0)$ & $(0, 0, 0)$ \\
\hline
\end{tabular}
\end{center}
Since the initial configuration has $(0, 1, 1)$ and the target $(0, 0, 1)$, the function application sequence with which the system produces the target $(0, 0, 1)$ must apply the first and second an indefinite number times in any order and then the third any positive number of times.

The effect of the third function on the last $k-1$ groups is to clear their states.
While applying the first and second groups, each of the $k$ groups has the result of applying the $k$ function groups of $\syst{F}$ in some order.
Specifically, for each $j \in [n]$, the $j$-th group has the result of applying a sequence ending with the $j$-th function group of $\syst{F}$.
Suppose we wish to apply the sequence with function group sequence $[j_1, \ldots, j_m]$, we let $j_0 = j_{m+1} = 1$ and execute $m$ rounds.
\begin{itemize}
\item
For a round $r \in [m]$, we let $\delta = j_{r} - j_{r-1}$ and adjust it to $\delta + k$ in the case where $\delta < 0$, execute the second group $\delta$ times, and then apply the first group.
\item
For round $m+1$, we execute shifting in the same manner as the previous rounds and then apply the thrdr group.
\end{itemize}
These actions allow to generate the result of applying the functions with index sequence $[j_1, \ldots, j_r]$ for each $r$ and then moving the final result to the first node group.

Since shifting is not necessary when $j_m = j_{m-1}$, the number of time steps $\syst{G}$ executes in one round is at most $(k-1) + 1 = k$.
Thus, to simulate $t$ steps of $\syst{F}$, $\syst{G}$ needs at most $k(t+1)$.
The ratio is $k(t+1) / t = k + 1/t \leq k + 1$.

In the case where $\syst{F}$ uses a permutation $\pi = [p_1, \ldots, p_n]$, we use a permutation
\[
p_1, \ldots, p_n, n + p_1, \ldots, n + p_n, \ldots, k(n-1) + p_1, \ldots, k(n-1) + p_n, kn + 1, kn + 2, kn + 3.
\]
We also change the functions for $y_1, y_2$, and $y_3$ as follows:
\begin{itemize}
\item
The first and second function groups copy $y_3$'s state to itself and to $y_1$ and copy $y_2$'s state to itself.
\item
The third function group copies $y_3$'s state to itself and to $y_2$ and copy $y_2$'s state to $y_1$.
\end{itemize}
 The initial states of $y_1, y_2$, and $y_3$ are $(0, 1, 0)$ and the target states of $y_1, y_2$, and $y_3$ are $(1, 0, 0)$.
The behavior of the three groups is as follows:
\begin{center}
\begin{tabular}{|c|c|c|c|}
\hline
& \multicolumn{3}{|c|}{input} \\
\cline{2-4}
& $(0, 1, 0)$ & $(1, 0, 0)$ & $(0, 0, 0)$ \\
\hline
Function group 1 & $(0, 1, 0)$ & $(0, 0, 0)$ & $(0, 0, 0)$ \\
\hline
Function group 2 & $(0, 1, 0)$ & $(0, 0, 0)$ & $(0, 0, 0)$ \\
\hline
Function group 3 & $(1, 0, 0)$ & $(0, 0, 0)$ & $(0, 0, 0)$ \\
\hline
\end{tabular}
\end{center}
This means that the target configuration is achievable only the first and second groups operate and then the third just once.
In the case of permutation list, we apply the same conversion to each permutation on the list, which preserves the number of permutations on the list.
\end{proof}

\section{Parallel or fixed-permutation updating combined with individual function selection}\label{sec:parallel-independent}

Regarding the upper bound for \problem{$t$-Reachability}, we have the following result.

\begin{theorem}\label{thm:parallel-independent}
The following are true for a $k$-choice, unary function DT-BFDS that selects functions individually for all the nodes and updates either in parallel or using one fixed sequence.
\begin{enumerate}
\item
For $k=2$, \problem{$t$-Reachability} is $\np$-complete if $t \geq 3$, $\nl$-complete if $t = 2$, and in $\ac^0$ if $t=1$.
\item
For $k=3$, \problem{$t$-Reachability} is $\np$-complete if $t \geq 2$ and in $\ac^0$ if $t=1$.
\item
For $k=2$, \problem{$t$-Predecessor} is $\np$-complete if $t \geq 2$ and $\nl$-complete if $t = 1$.
\item
For $k=3$, \problem{$t$-Predecessor} is $\np$-complete if $t \geq 1$.
\end{enumerate}
\end{theorem}

\begin{proof}~
[(1)]~
Here is the proof for $t=3$.
The proof uses a reduction from \problem{3SAT}.
Let $\phi = C_1 \wedge \cdots \wedge C_m$ be a 3CNF formula over $n$ variables $x_1, \ldots, x_n$.
We will construct a $2$-choice DT-BFDS as we describe next.

The system has four levels of nodes as we describe below.
\begin{itemize}
\item[(i)]
The first level consists of only two nodes, $a_0$ and $a_1$, constantly representing $0$ and $1$.
In other words, the update function is the identify function for both $a_0$ and $a_1$ in both function choices, and the initial configuration is $0$ for $a_0$ and $1$ for $a_1$.
\item[(ii)]
The second level has $n$ pairs of nodes, $(b_{i,0}, b_{i,1}), i \in [n]$.
These nodes are to represent a truth-assignment to the variables $x_1, \ldots, x_n$.
The node $b_{i,1}$ represents the positive literal of $x_i$.
The node $b_{i,0}$ represents the negative literal of $x_i$. 
For each $i \in [n]$, the two unary functions of $b_{i,0}$ and $b_{i,1}$ take input from $a_0$ and $a_1$.
\item[(iii)]
The third level has two parts.
One part has $n$ pairs, $(c_{i,0},c_{i,1}), i \in [n]$.
For each $i \in [n]$, the two unary functions of $c_{i,0}$ and $c_{i,1}$ take inputs from $b_{i,0}$ and $b_{i,1}$.
The other part has $2m$ nodes, $\alpha_j, \beta_j, j \in[m]$.
For all $j \in [m]$, the unary functions of $d_j$ take input from the first two literals of $C_j$ that the second level nodes represent.
For all $j \in [m]$, the unary functions of $d'_j$ take input from the last two literals of $C_j$ that the second level nodes represent.
For example, if $C_j = x_4 \lor \overline{x_5} \lor \overline{x_9}$, then $\alpha_j$'s two unary functions take input respectively from $b_{4,1}$ and $b_{5,0}$, and $\beta_j$'s two unary functions take input respectively from $b_{5,0}$ and $b_{9,0}$.
\item[(iv)]
The fourth level has two parts.
The first part has $n$ pairs, $(d_{i,0},d_{i,1}), i \in [n]$.
For each $i \in [n]$, both unary functions of $d_{i,0}$ take input from $c_{i,0}$ and both unary functions of $d_{i,1}$ take input from $d _{i,1}$.
The second part has $m$ nodes, $\gamma_j, j \in[m]$.
For each $j \in [m]$, $\gamma_j$'s unary functions take input from $\alpha_j$ and $\beta_j$.
\end{itemize}
In the initial configuration $\config{c}$, every node's state is $0$ except $a_1$'s state is $1$.
In the target configuration $\config{d}$, $\gamma_j$'s state is $1$ for all $j \in [m]$, $a_1$'s state is $1$, $e_{i,1}$'s state is $1$ for all $i \in [n]$, and all other nodes are $0$.

Every path from a node on the first level to a node on the second level has length $1$,
every path from a node on the first level to a node on the third level has length $2$, and
every path from a node on the first level to a node on the fourth level has length $3$.
There is no feed-back loop except for the self-loops at $a_{0}$ and $a_1$.
Let $t\geq 3$.
Then the configuration after the $t$-th time step is dependent only on the action that the system chooses at the $(t-1)$-st, $(t-2)$-nd, and $(t-3)$-rd time steps.
More precisely,
\begin{itemize}
\item
The states of the nodes on the fourth level after time step $t$ depend on:
\begin{itemize}
\item
the system's action on the fourth level at time step $t$, 
\item
the system's action on the third level at time step $t-1$, and
\item
the systems action on the second level at time step $t-2$.
\end{itemize}
\item
The states of the nodes on the third level after time step $t$ depend on:
\begin{itemize}
\item
the system's action on the third level at time step $t$ and
\item
the system's action on the second level at time step $t-1$.
\end{itemize}
\item
The states of the nodes on the second level after time step $t$ depend on:
\begin{itemize}
\item
the system's action on the second level at time step $t$.
\end{itemize}
\end{itemize}
In light of this observation, we have that for the desired pattern to appear on $d$'s, $c$'s, and $b$'s:
\begin{enumerate}
\item
For all $i \in [n]$, at time step $t-2$, the system must copy $1$ to one of $b_{i, 0}$ and $b_{i, 1}$ and $0$ to the other.
\item
For all $i \in [n]$, at time step $t-1$, the system must copy the $1$ from $b_{i, 0}$ and $b_{i, 1}$ to $c_{i, 1}$ and the $0$ to $c_{i, 0}$, for  and $0$ to the other.
\item
For all $i \in [n]$, at time step $t-1$, the system must copy the $0$ from $a_0$ to both $b_{i, 0}$ and $b_{i, 1}$.
\item
For all $i \in [n]$, at time step $t-1$, the system will copy the $0$ from either $b_{i, 0}$ or $b_{i, 1}$ to $c_{i, 0}$ and $c_{i, 1}$.
\item
For all $i \in [n]$, at time step $t$, the system will copy the $0$ from $c_{i, 0}$ to $d_{i, 0}$ and the $1$ from $c_{i, 1}$ to $d_{i, 1}$.
\end{enumerate}
Also, we have:
\begin{enumerate}
\item
For all $j \in [m]$, at time step $t$, the system must copy $1$ from either $\alpha_j$ or $\beta_j$ to $\gamma_j$.
\item
For all $j \in [m]$, at time step $t-1$, the system must copy $1$ to one of $\alpha_j$ or $\beta_j$.
\item
For all $i \in [n]$, at time step $t-2$, the system must copy $1$ to one of the $b$'s representing the literals of $C_j$.
\item
For all $j \in [m]$, at time step $t$, the system will copy $0$ to both $\alpha_j$ and $\beta_j$.
\item
For all $i \in [m]$, at time step $t-1$, the system will copy the $0$ from $a_0$ to the $b$'s representing the literals of $C_j$.
\end{enumerate}
The first requirement is the same as saying that the assignment the system generates at time step $t-2$ is a satisfying assignment.
The last requirement is identical to the third requirement from the previous list.
Thus, $\config{d}$ is reachable after time step $t$ if and only if $\phi$ is satisfiable.
We thus have that \problem{$t$-Reachability} is $\np$-hard.
The membership in $\np$ follows from the general upper bound result.

In the case of $t = 2$, suppose that the $\phi$ in the aforementioned construction is a 2CNF formula.
We can follow the same construction where the $\alpha$'s act in place of the $\gamma$'s and $d$'s act in place of the $c$'.
Then, $\phi$ is satisfying if and only if $\config{d}$ is reachable in two time steps.
The proof of the two time-step reachability in $\nl$ uses the following logic.
Suppose we have a system $\syst{F}$ and two configurations $\config{c}$ and $\config{d}$ and must decide if $\config{d}$ is reachable from $\config{c}$ in two time steps.
Suppose $\syst{F}$ has $n$ nodes.
Let $x_1, \ldots, x_n$ be variables representing the states after the first time step.
Suppose the $i$-th node takes input from either the $k$-th node or the $\ell$-th node.
We introduce the following 2-literal clause depending of the $d_i$, the $j$-th node's state in $\config{d}$:
\begin{center}
$(x_k \vee x_{\ell})$ if $d_i = 1$ and $\overline{x_k} \vee \overline{x_{\ell}}$ otherwise.
\end{center}
Also, we introduce the following 1-literal clause depending on $c_k$ and $c_{\ell}$:
\begin{center}
$(x_k)$ if $c_k = c_{\ell} = 1$ and
$(\overline{x_k})$ if $c_k = c_{\ell} = 0$.
\end{center}
Let $\phi$ be the conjunction of these clauses.
Clearly, $\phi$ is a 2CNF formula.
A satisfying assignment of $\phi$ represent the intermediate state values on one path from $\config{c}$ to $\config{d}$.
A logarithmic-space machine can compute the formula by scanning the encoding of $\syst{F}$ and the two configurations.
Thus, the reachability is in $\nl$-complete.

In the case of $t=1$, the $x$'s in the previous proof are the entries of the configuration $\config{c}$.
So, testing can be by way of the AND of $n$ fan-in-$2$ ORs.
Thus, it is in $\ac^0$.

\medskip

[(2)]~
In the case where $k=3$, the use of two node groups, $\alpha$'s and $\beta$'s, is not necessary.  The nodes $\gamma$'s can directly select values from the nodes that represent their literals.
This means that two time steps will be sufficient for testing the satisfiability of a CNF formula.
The testability in $\ac^0$ still holds.

\medskip

[(3)]~
The proof for the predecessor existence problem follows the same proof as (1).
We try to move back from $\mathbf{d}$ to $\mathbf{c}$.
Since each pair of $d$'s has one $0$ and one $1$ and the pair $a$ has one $0$ and $1$, after three backward steps, if possible at all, each pair of $c$'s, $b$'s, and $a$ has one $0$ and $1$.
The $1$'s appearing in the $\gamma$ part should be trace-able back to the $a$'s.
From these, we can see that the system can go back for three steps, in the case where the formula $\phi$ is satisfiable.
In fact, the three-step backward traversal can start at any point.
Thus, a $t$-predecessor exists for all $t\geq 2$.
In the case where $\phi$ is not satisfiable, regardless of how the system chooses the values of $\alpha$'s and $\gamma$'s, going back one more step is impossible in a manner consistent with the requirement that $b$'s have one $0$ and one $1$ for each pair, and so $t$-predecessors do not exist for $t = 2$.

For the $\nl$-completeness proof, we think of the reduction from a 2CNF formula as in part (1).
In that proof, we directly went from the truth-assignment at $b$'s level to $\gamma$'s. Using the same course of argument, we see that the $1$-predecessor problem is $
nl$-complete.

\medskip

[(4)]~The proof uses the same idea for the $\nl$-completeness in the previous part. Since $k=3$, there are three choices for $\gamma$, and so we can move up from 2CNF to 3CNF.
\end{proof}

Given the $\np$-completeness, we naturally wonder if the reduction is usable to show the $\sharpp$-completeness of the corresponding counting problem. 
A many-one reduction $f$ from an $\np$-language $A$ to some other $\np$-language $B$ is a witness-preserving reduction~\cite{gar-joh:b:np} if $f$ has the following property:
Concerning some witness schemes for $A$ and $B$,  for each $x\in A$, the number of witnesses for $x$ is $K(x)$ times the number of witnesses for $f(x)$ for some $K(x)$, and each witness for $x$ corresponds to exactly $K(x)$ witnesses for $f(x)$.
Thus, if an $\np$-complete problem has a witness-preserving reduction from \problem{SAT}, the reduction naturally indices $\sharpp$-completeness of the corresponding counting problem.

If four choices are allowed, we can observe that a modification of a proof for Theorem~\ref{thm:parallel-independent} gives a witness-preserving reduction where $K$ is the identity function.
\begin{corollary}\label{coro:parallel-independent_counting}
	\problem{Counting Path} is $\sharpp$-hard under parsimonious reduction for a $4$-choice, unary function DT-BFDS that selects functions individually for all the nodes and updates either in parallel or using one fixed sequence.
\end{corollary}
\begin{proof}
	Recall the proof for Theorem~\ref{thm:parallel-independent} where we reduce 3SAT to \problem{$t$-Reachability}. In the reduction, we are given a 3CNF formula $\phi = C_1 \wedge C_2 \wedge \cdots \wedge C_m$, and construct a 3-choice DT-BFDS consisting of nodes $a_0, a_1$, $b_{i, 0}, b_{i, 1}, c_{i, 0}, c_{i, 1}$, $i \in [n]$, and $\gamma_j$, $j \in [m]$. We employ slightly different update functions defined as follows:
	\begin{itemize}
		\item The node $a_0$ has only one update function that is the identity function taking input from $a_0$ itself;
		\item The node $a_1$ has only one update function that is the unary function taking input from $a_0$;
		\item For every $i \in [n]$, each of $b_{i, 0}$ and $b_{i, 1}$ have two update functions. One takes input from $a_0$, and the other takes input from $a_1$;
		\item For every $i \in [n]$, $c_{i, 0}$ has two update functions. One takes input from $b_{i, 0}$, and the other takes input from $b_{i, 1}$;
		\item For every $i \in [n]$, $c_{i, 1}$ has three update functions. One takes input from $b_{i, 0}$, another takes input from $b_{i, 1}$, and the other is the identity function that takes input from $c_{i, 1}$ itself; and 
		\item For every $j \in [m]$, $\gamma_j$ has four update functions. Three unary functions take input from the literals of $C_j$ that the second level nodes represent. For example, if $C_j = x_4 \lor \overline{x_5} \lor \overline{x_9}$, then the three unary functions take input respectively from $b_{4,1}$, $b_{5,0}$ and $b_{9,0}$. The other function is the identity function that takes input from $\gamma_j$ itself.
	\end{itemize}
	Let $F$ be the resulting DT-BFDS. Clearly, $F$ is a 4-choice DT-BFDS. In the initial configuration $\vec{c}$, every node's state is $0$ except $a_1$'s state is $1$. We employ a slightly different target configuration $\vec{d}$ from the one in Theorem~\ref{thm:parallel-independent}, where $\gamma_j$'s state is $1$ for all $j \in [m]$, $c_{i,1}$'s state is $1$ for all $i \in [n]$, and all other nodes are $0$ (That is, $a_1$ has the value 0).
	
	Note that only $a_i$ contains the value 1 among the nodes in $\vec{c}$. The value 1 in $a_1$ is propagated to some of $b_{i,0}$ and $b_{i,1}$ in the first round, and then to some of $c_{i, 1}$ and $\gamma_j$ in the second round. Since no backward directions are allowed, $c_{i, 1}$ and $\gamma_j$ need to keep the values to reach $\vec{d}$, while the values in $a_0, a_1, b_{i, 0}$ and $b_{i, 1}$ are 0s in the following rounds. Thus, every path from $\vec{c}$ to $\vec{d}$ in the configuration graph is of lenght 2 with self-loops at $\vec{d}$. Therefore, the number of simple paths from $\vec{c}$ to $\vec{d}$ is the number of paths of length two from $\vec{c}$ to $\vec{d}$.
	
	We now count the paths of length two in the configuration graph, and show that it is equal to the number of the satisfying assignments of $\phi$. We can observe that, for each satisfying assignment of $\phi$, there exists a unique configuration $\vec{e}$ such that $F(\vec{c}) = \vec{e}$ and $F(\vec{e}) = \vec{d}$, as follows. Let $\xi = (\xi_1, \xi_2, \dots , \xi_n) \in \{ 0, 1\}^n$ be an arbitrary satisfying assignment of $\phi$. The unique configuration $\vec{e}$ has the following values in the nodes:
	\begin{itemize}
		\item $a_0$'s state is 0 and $a_1$'s state is 1;
		\item For every $i \in [n]$, $b_{i,0}$'s state is 1 if and only if $\xi_i = 0$, and $b_{i,1}$'s state is 1 if and only if $\xi_i = 1$;
		\item For every $i \in [n]$, both $c_{i, 0}$'s state and $c_{i, 1}$'s state are 0s; and
		\item For every $j \in [m]$, $\gamma_j$'s state is 0.
	\end{itemize}
	Note that the values for $a_0, a_1$, $c_{i, 0}$s and $j \in [m]$, $\gamma_j$s in the first round are constant no matter how individual choices are made. Furthermore, the values in $b_{i,0}$ and $b_{i,1}$ are determined by $\xi$. Thus, the number of paths of length two is equal to the number of the satisfying assignments of $\phi$.
\end{proof}

With the individual function selection scheme, a unary function can choose between $0$ and $1$, where one source is $1$, and another is $0$.
The availability of both positive and negative unary functions allows nondeterministic selections between $0$ and $1$ from a single source.
Does the negative unary function add more computational power?
We have the following partial answer that utilizes the doubling trick with bits flipped on the copy.

\begin{theorem}\label{thm:negation}
Let $\syst{F}$ be an $(n,k)$ system with an arbitrary function selection scheme.
Suppose each function of $\syst{F}$ is unary.
Suppose the update scheme of $\syst{F}$ is parallel, fixed permutation, or permutation list.
Then there is a $(2n, k)$ system using only the positive unary function and the same updating scheme that can simulate $\syst{F}$.
\end{theorem}

\begin{proof}
Let $n$, $k$, and $\syst{F} = \{ f_{i,j} \mid i \in [n], j \in [k] \}$ be as in the statement of the theorem.
Let $x_1, \ldots, x_n$ be the variables representing the nodes' states for $\syst{F}$.
We will construct a new system $\syst{G} = \{ g_{i, j} \mid i \in [2n], j \in [k] \}$ with $2n$ nodes, where we embed a configuration $\config{c}$ of $\syst{F}$ as $\config{c} \config{c}'$, where $\config{c}'$ is the position-wise negation of $\config{c}$.
In other words, for $\config{c} = (c_1, \ldots, c_n)$, $\config{c}' = (\overline{c_1}, \ldots, \overline{c_n})$.
Let $y_1, \ldots, y_{2n}$ be the variables representing the nodes' states for $\syst{G}$.
The definition of $g_{i,j}$'s is as follows:
\begin{itemize}
\item
If $f_{i,j} = x_\ell$, then $g_{i, j} = y_\ell$ and $g_{i + n, j} = y_{\ell + n}$.
\item
If $f_{i,j} = \overline{x_\ell}$, then $g_{i,j} = y_{\ell + n}$ and $g_{i + n, j} = y_{\ell}$.
\end{itemize}
In the case of parallel updating, $g_{i, j}$ and $g_{i+n, j}$ work as a pair that computes $f_{i, j}$ and its complement.
In the case of fixed-permutation updating and permutation-list updating, each permutation $\pi$ becomes the permutation $\sigma$:
\[
\pi(1), \pi(1) + n, 
\pi(2), \pi(2) + n,
\cdots, 
\pi(n), \pi(n) + n. 
\]
Since for all $i \in [n]$, $g_{i, j}$ and $g_{i + n, j}$ do not depend on each other, the sequence works as the same as $\pi$.
\end{proof}

Is it possible to reduce the number of nodes endowed with function choices while preserving the computational complexity of the models? If so, how much?
We have obtained a partial answer to the question.

\begin{theorem}\label{thm:four-nodes-unary}
For each $n \geq 4$, there exists a unary $(n, 2)$ DT-BFDS $\syst{F}$ that updates in parallel and makes individual function selections such that $n - 2$ nodes have two identical function choices and the configuration graph is fully connected, except for $0^n$ and $1^n$.
\end{theorem}

\begin{proof}
Let $n\geq 4$.
We define a system with $n$ nodes with the state variables $x_1, \ldots, x_n$ with the following update functions $f_{1, 1}, \ldots, f_{n, 1}, f_{1, 2}, \ldots, f_{n, 2}$.
\begin{itemize}
\item
For each $i, 3\leq i\leq n$, $f_{i, 1} = f_{i, 2} = x_{i-1}$.
\item
$f_{1, 1} = x_1$ and $f_{1, 2} = x_n$.
\item
$f_{2, 1} = x_n$ and $f_{2, 2} = x_1$.
\end{itemize}
Suppose the system uses $f_{i,1}$ for all $i$.
Since $f_{2,1} = x_n$ and $f_{1,1} = x_1$, the first group of functions works as the rotation among $[x_2, \ldots, x_n]$ and one among $x_1$.
The first group thus preserves the number of $0$s and the number of $1$s appearing as the states.
If $x_1$ uses $f_{1, 2}$ instead, the value $x_1$ is copied to $x_1$.
Also, if $x_2$ uses $f_{2, 2}$ instead, the state in $x_1$ is copied to $x_2$.
In this manner, if a state in the cycle $[x_2, \ldots, x_n]$ need a change, say from $0$ to $1$, and there is already one $1$ appearing in the cycle, we can do the following:
\begin{itemize}
\item
Keep rotating.
\item
Copy one of the $1$s to $x_1$ while a $1$ appears at $x_n$ while rotating the $n-1$ states.
\item
Keep rotating.
\item
Copy the $1$s at $x_1$ to the $0$ where the $0$ appears at $x_n$.
\end{itemize}
This method fails to work where the cycle has no $1$s and $x_1 = 0$.
If the cycle has no $1$s and $x_1 = 1$, we can assume that the first two steps are complete.
The same argument holds with the $0$ and $1$ switching their roles.
This means that $0^n$ and $1^n$ are cycles in the configuration graph and the others are fully connected.
This proves the theorem.
\end{proof}

We do not know if the number can be smaller than $2$ (that is, equal to $1$).

\section{Parallel updates where the function selections are coordinated}\label{sec:coordinated}

For each $k\geq 1$, we can assume that the dependency graph of a $k$-choice DT-BFDS has $k$ layers, and the $k$ layers have a one-layer, collapsed representation.
If a constant $d$ bounds the maximum fan-in , then the in-degree of the aggregate graph is at most $d k$.

As stated earlier, \problem{Reachability} is $\pspace$-complete for the deterministic model if the permissible function types form a complete Boolean basis.
In light of the result, we question if restricting the update functions' computational power will result in the characterization of a class below $\pspace$.
The most restricted dependency structure is where every node has in-degree $1$ and out-degree $1$.
The degree constraints make each dependency graph disjoint directed cycles.
Within each cycle, the state values rotate among the nodes in it; that is, for a cycle $w = [w_1, \ldots, w_m]$ with the state values $[s_1, \ldots, s_m]$, applying the cyclic changes turns the state values to $[s_m, s_1, \ldots, s_{m-1}]$.
The number of $1$s and $0$s in the state values are unchanged during the operation.

\begin{definition}\label{def:cyclic-systems}
\deftitle{Cyclic and pure cyclic systems}
An $(n, k)$ DT-BFDS $\syst{F} = \{ f_{i, j} \mid i \in [n], j\in [k] \}$ is \myemph{cyclic} for all $j \in [k]$, the dependency graph induced by the $j$-th group $(f_{1,j}, \ldots, f_{n,j})$ is a collection of independent simple directed cycles.
The functions of $\syst{F}$ are all unary functions (positive or negative).
In addition, we say that the system is \myemph{pure cyclic} if all the functions are positive unary functions.
\end{definition}

The idea of using cyclic BFDS is reminiscent to~\cite{ogi-uch:c:synchronous}, which shows that \problem{Reachability} and \problem{Predecessor} for these models are generally easy to solve (even in $\nc$).
(The paper~\cite{ogi-uch:c:synchronous} uses the term ``permutational'' for ``cyclic.'')

In the case of $k$-choice, pure cyclic DT-BFDS that update in parallel and select functions in coordination, we question how difficult it is to determine the properties of the configuration graph?
We have a partial answer to the question.
The \problem{Graph Isomorphism with Basis} is the problem to answer, given two graphs $G$ and $H$ having some $n$ nodes and base permutations $\pi_1, \ldots, \pi_k$ over $[n]$, $G$ is transformable to $H$.

\begin{theorem}\label{thm:isomorphism-basis}
For parallel $2$-choice, pure cyclic DT-BFDS that selects functions in coordination, \problem{Graph Isomorphism with Basis} is polynomial-time many-one reducible to \problem{Reachability}.
\end{theorem}

\begin{proof}
Let $G = (V, E)$ be a graph with $n$ nodes for some $n\geq 1$.
Let $V = \{ v_1, \ldots, v_n \}$.
Let $H = (V, D)$ be another graph having the same number of nodes.
We need to test $G$ and $H$ are isomorphic to each other.
We construct an $n^2$-node system $\syst{F}$.
Use $\config{c}$ be a configuration of $\syst{F}$ representing the graph $G$'s adjacency matrix $(e_{ij}), i, j \in [n]$, where $e_{ij} = 1$ if $i\neq j$ and there is an edge between the $i$-th node the $j$-node and $0$ otherwise.
Let $\tau(i, j) = i(n-1) + j$.
The function $\tau$ is one that provides the mapping from the double indexing for $E$ to the single indexing of $\syst{F}$.
Let $\config{d}$ be the one for the graph $H$.
We define two function groups $R = \{ r_1, \ldots, r_{n^2} \}$ and $S = \{ s_1, \ldots, s_{n^2} \}$.
The first group $R$ acts as the rotation among the nodes $v_2, \ldots, v_n$ such that the rotation moves $v_i$ to $v_{i-1}$ for all $i, 3\leq i\leq n $, and moves $v_2$ to $v_n$.
The group $R$ does not move $v_1$.
Let
\[
\alpha(i) = \left\{ \begin{array}{ll} 1 & \mathrm{if~}i=1\\
i+1 & \mathrm{if~}2\leq i\leq n-1\\
2 & \mathrm{othewise}\\
\end{array} \right. 
\]
The function group $R$ is then
\[
r_{\tau(i, j)} = x_{\tau( \alpha(i), \alpha(j) }.
\]
We similarly define the group $S$ using the rotation between $v_1$ and $v_2$, which keeps everything else thesame.
Let
\[
\beta(i) = \left\{ \begin{array}{ll} 2 & \mathrm{if~}i=1\\
1 & \mathrm{if~}i = 2\\
i & \mathrm{othewise}\\
\end{array} \right. 
\]
The function group $R$ is then
\[
s_{\tau(i, j)} = x_{\tau( \beta(i), \beta(j) }.
\]
The two rotations form a two-element basis for the permutations over $[n]$.
Thus, $G$ and $H$ are isomorphic to each other if and only if $\config{c}$ if transformable to $\config{d}$ using $R$ and $S$.
Since the functions appearing $R$ and $S$ are all unary functions, the claim holds.
\end{proof}

Noting that the basis from the proof has the following property.
First, by combining at most $2n+1$ of them, we can construct a permutation between any neighboring indices, $i$ and $i+1$ (we will treat $n+1$ as $1$).
Next, using at most $2n+1$ neighboring exchanges, we can construct the permutation between any $i$ and $j$, $i\neq j$.
Finally, by combining at most $n$ of these, we can construct any permutation.
The total number of applications of the basis function is at most
\[
(2n+1)^2n = 4n^3 + 4n^2 + n \leq 9n^3 = O(n^3). 
\]
This gives the following corollary.

\begin{corollary}\label{coro:graph-iso}
\problem{Graph Isomorphism} is polynomial-time many-one reducible to \problem{Reachability} for a $2$-choice, pure cyclic DT-BFDS that updates in parallel and selects functions in coordination.
The reduction has a property: if the answer to the reachability question is positive, there is a path having length $d n^{\alpha}$ for some constants $d$ and $\alpha$.
\end{corollary}

It is not hard to show that there is a $2$-choice cyclic DT-BFDS in which the configuration graph is strongly connected. In other words, for all pairs of configurations, $(\config{c}, \config{d})$, $\config{d}$ is reachable from $\config{c}$.

\begin{proposition}\label{prop:coordinated-aprallel}
Let $\syst{F}$ be a $2$-choice, cyclic DT-BFDS that updates in parallel and selects functions in coordination such that (i) the functions $(f_{1,1}, \ldots, f_{n,1})$ induce one single directed simple cycle and (ii) the functions $f_{1,2}, \ldots, f_{n,2}$ are all self-loops with $f_{1,2}$ is the negation of itself.
Then, for all configurations $\config{c}$ and $\config{d}$, $\config{d}$ is reachable from $\config{c}$.
\end{proposition}

\begin{proof}
The result immediately follows from the discussion in the proof of Theorem~\ref{thm:isomorphism-basis}.
\end{proof}

Noting that \problem{Graph Isomorphism} is in $\np$, we ask if \problem{Reachability} becomes $\np$-complete in some settings.
We have a partial answer to the question.

\begin{theorem}\label{thm:coordinated-npc}
If $t$'s specification is in unary and the function basis consists of the $2$-fan-in OR and the unary, then \problem{$t$-Reachability} for $4$-choice DT-BFDS that update in parallel and select functions in coordination is $\np$-complete. 
\end{theorem}

\begin{proof}
Let $\phi = C_1 \wedge \cdots \wedge C_m$ be a 3CNF formula over $n$ variables $x_1, \ldots, x_n$.
We will construct a system $\syst{F}$.
The system's nodes are in four parts.
\begin{itemize}
\item
The \emph{assignment gadget} consists of $2(n+1)$ nodes, $a_0, \ldots, a_n, b_0, \ldots, b_n$.
\item
The \emph{evaluation gadget} consists of $3m$ nodes, $\alpha_i, \beta_i, \gamma_i, i\in[m]$.
\item
The \emph{flow checking gadget} consists of $(n+2)^2+3$ nodes.
The are $c_{i,j}, 0\leq i\leq n+1, 0\leq j\leq n+1$, and $d_1, d_2, d_3$.
\item
The \emph{step counting gadget} consists of $t_0, \ldots, t_{2n+4}$.
\end{itemize}
The initial configuration is all $0$ except for $a_0 = c_{1, 1} = t_1 = 1$.
The target configuration has $1$ at all $a$'s, $b$'s, $\alpha$'s, $\beta$'s, $\gamma$'s, $c_{n+1,n+1}$, $d_1, d_2$, $d_3$, and $t_{2n+4}$; the remaining nodes have $0$.

The first function group acts on $a_0, b_0$, the flow checking gadget, and the step counting gadet.
On $a_0$ and $b_0$, the group exchanges the states between $a_0$ and $b_0$, going back and forth between $(0,1)$ and $(1,0)$.
On the action checking gadget, concurrently for each $i, 1\leq i\leq n+1$, the group copies the state from $c_{i-1,j}$ to $c_{i,j}$ for all $j, 0\leq j\leq n+1$.
On the step counting gadget, the group copies the state from $t_i$ to $t_{i+1}$ for all $i, 0\leq i\leq 2n+3$.
The group preserves the state of all other nodes.

The second function group acts on $a_i, b_i, i\in[n]$, the flow checking gadget, and the step counting gadget.
Concurrently for each $i \in [n]$, the group copies the state of $a_{i-1}$ to $a_i$ and the state of $b_{i-1}$ to $b_i$.
On the flow checking gadget, concurrently for each $j, 1\leq j\leq n+1$, the group copies the state from $c_{i,j-1}$ to $c_{i,j}$ for all $i, 0\leq j\leq n+1$.
On the step counting gadget, the group copies the state from $t_i$ to $t_{i+1}$ for all $i, 0\leq i\leq 2n+3$.
The group preserves the state of all other nodes.

The third group acts on $\alpha_i$ and $\beta_i$, $i \in [m]$, the flow checking gadget, and the step counting gadget.
The group treats $a_1, \ldots, a_n$ as the negative literals corresponding to the variables $x_1, \ldots, x_n$ and $b_1, \ldots, b_n$ as the positive literals corresponding to the variables $x_1, \ldots, x_n$.
With the treatment, the group stores the OR of the first two literals of $C_i$ to $\alpha_i$ and the OR of the last two literals of $C_i$ to $\beta_i$ for each $i \in [m]$.
On the flow checking gadget, it copies the state from $c_{n+1,n+1}$ to $d_1$ and from $d_2$ to $d_3$.
On the step counting gadget, the group copies the state from $t_i$ to $t_{i+1}$ for all $i, 0\leq i\leq 2n+3$.
The group preserves the state of all other nodes.

The last group acts on $\gamma_i, i \in [n]$, $a_i, b_i, 0\leq i\leq n$, the flow checking gadget, and the step counting gadget.
For each $i\in [n]$, the group stores the OR of $\alpha_i$ and $\beta_i$ in $\gamma_i$.
Also, for each $i, 0\leq i\leq n$, the group stores the OR of $a_i$ and $b_i$ to both $a_i$ and $b_i$.
On the flow checking gadget, it copies the state from $d_1$ to $d_2$.
On the step counting gadget, the group copies the state from $t_i$ to $t_{i+1}$ for all $i, 0\leq i\leq 2n+3$.
The group preserves the state of all other nodes.

We claim that $\phi$ is satisfiable if and only if the target configuration is reachable from the initial configuration, and furthermore that if the target is reachable, then the target emerges precisely after $2n+3$ time steps and disappears with one more time step.

First, the action on the step counting gadget is the same for all the function groups.
If there is $1$ appearing on the gadget, the action is to move the one to the next position.
If the $1$ is at $t_{2n+4}$, any one of the function removes the $1$ because $t_{2n+3}$ has a $0$.
Thus, achieving the target configuration on the step counting gadget requires exactly $2n+3$ total applications of the function groups.

The third and the last function groups act on $d_1, d_2, d_3$.
For them to be able to take in a $1$, the $1$ initially located at $c_{1,1}$ must be brought to $c_{n+1, n+1}$.
The point $c_{n+1,n+1}$ has Manhattan distance $2n$ from $c_{1,1}$.
So, the total number of applications of the first two groups must be $2n$.
The first group makes the horizontal shifts and ``Column'' $n+1$ is the boundary, the first group occurs exactly $2n$ times.
This means that the second group occurs exactly $n$ times.
Setting the values of $d_1, d_2, d_3$ to $1$ in $3$ time steps is only accomplishable by applying the third group, the fourth group, and the third group in this order.

The first group swaps the states between $a_0$ and $a_1$.
The second group copies $a_0, \ldots, a_{n-1}$ to $a_1, \ldots, a_n$, respectively, and does the same for the $b$'s.
Thus, when the third group occurs for the first time, exactly one of $a_i$ and $b_i$ is  $1$ for all $i, 0\leq i\leq n$.
We can think the value pair $(a_i, b_i)$ at the time step to be a truth-assignment to the variables of $\phi$.
We then can think of the application of the third group and the fourth group in succession as the evaluation of $\phi$ with the truth-assignment.
Since the fourth group acts only once, the $\gamma_1 = \cdots = \gamma_m$ in the target configuration achievable if and only if the truth-assignment that $a$'s and $b$'s represent is a satisfying assignment.

The last application of the third group turns all $a$'s, $b$'s, $\alpha$'s and $\gamma$'s to $1$.
Since the pair $(a_0, b_0)$ is either $(0, 1)$ or $(1, 0)$, prior to an application of the second group, one or zero applications of the first group is necessary.
More precisely, the $j$-th application of the second group is for selecting the truth-assignment for $(\overline{x_{n+1-j}}, x_{n+1-j})$.
For $(\overline{x_{n}}, x_{n})$, the first group may or may not be necessary.
For $(\overline{x_{n+1-j}}, x_{n+1-j})$, one application is necessary if the assignment is opposite to the previous one and is unnecessary otherwise.
The total number of applications of the first group when the truth-assignment is complete can be less than $n$.
If that is the case, the first group occurs the number of times equal to the difference.

This proves the theorem.
\end{proof}

Another interesting question is the cycle length.
In the case where $k=1$, the pure/cyclic $k$-choice model is so simple that \problem{Reachability} is in polynomial time~\cite{ogi-uch:c:synchronous}.

\section{Permutation-list/Arbitrary-permutation models}\label{sec:nondeterministic-sequential-updating}

\subsection{$2$-choice models that select functions independently}\label{sub:2-choice}

A good starting point for studying models with multiple possible updating sequences is the model with deterministic function choices.

An interesting question is under what conditions the reachability problem becomes $\np$-complete.
We have the following result.

\begin{theorem}\label{thm:sequential-individual-time-2}
\problem{$t$-Reachability} for a $2$-choice DT-BFDS that uses a permutation list is $\np$-complete if $t=2$ and in $\ac^0$ if $t=1$.
\end{theorem}

\begin{proof}
When $t=2$, we can think of two different permutations may act in the two time steps.
Recall the proof for Part (1) of Theorem~{thm:parallel-independent}.
Let $\pi$ be a permutation in which the actions occur in the order of:
\begin{center}
$a$'s, $b$'s, $c$'s, $d$'s, $\alpha$'s, $\beta$'s, and then $\gamma$'s.
\end{center}
Let $\sigma$ be a permutation in which the actions occur in the order of:
\begin{center}
$\gamma$'s, $a$'s, $a$'s, $b$'s, $c$'s, $d$'s, $\alpha$'s, and then $\beta$'s.
\end{center}
If the input format $\phi$ is satisfiable, it is possible to produce the target configuration in two time steps with $\pi$ going first and then $\sigma$ going second.
If the 
\end{proof}

For \problem{$1$-Reachability} with the permutation-list or the arbitrary-permutation schedule, the problem of counting updating sequences that take the initial configuration to the target configuration is in $\sharpp$.
Can the problem be $\sharpp$-complete?

\subsection{Models with multiple choices for functions and sequences}\label{sub:robustness}

The next step in the project is to investigate the computational power of sequential DT-BFDS where uncertainty exists for updating schedule and update function selection.

\myemph{$3$-Reachability} is $\np$-complete for multiple-choice systems that use individual function selections, and \myemph{$2$-Reachability} is $\np$-complete for systems with a fixed permutation.
In~\cite{ara-fan-etal:k:combinatorics}, the authors show that it is $\np$-complete to decide, given ordering constraints such as ``$a$ must come before $b$'', whether there is an ordering satisfying all the constraints.
It is possible to restate \problem{$1$-Reachability} as a similar problem.
However, the constraints come from the two configurations.
It is unknown if the problem is still $\np$-complete.

\boxedq{q:sequence-existence-1}{
\myemph{($1$-Reachability)~}
Is the problem $\np$-complete for multiple-choice, multiple-permutation systems?
}

The models with function choices and sequences choices have an alternate interpretation.
The interpretation is a $2$-player game, where one player determines the updating sequence at each time step, and the other chooses the functions to execute.
Suppose that after receiving an initial configuration, the second player's goal is to force the system to arrive at the target configuration, while the first player's goal is to prevent the system from arriving at the target configuration. 

\begin{definition}\label{def:robustness}
\deftitle{Robustness}
Let $n, k \geq 1$.
Let $\syst{F}$ be a sequential $(n, k)$ DT-BFDS that selects functions individually for all the nodes and executes updates using either a sequence list or arbitrary sequences.
Let $t \geq 0$.
Then, $\syst{F}$ \myemph{robustly takes} $\config{c}$ to $\config{d}$ in $t$ time steps if the following property holds.
If $t=0$, $\config{c} = \config{d}$;
if $t \geq 1$: for every permissible sequence $\pi$, there exists a function selection $J$ and exists a configuration $\config{c}'$ $\syst{F}_{\pi}[J](\config{c}) = \config{c'}$ and $\syst{F}$ robustly takes $\config{c'}$ and $\config{d}$ in $t-1$ time steps.
\end{definition}

Inspiration for the above robustness computation comes from alternation~\cite{cha-koz-sto:j:alternation} and~\cite{hem-ogi:j:universally-serializable} about robust computation over a monoid.
A straightforward complexity upper bound of \problem{Robust $t$-Reachability} is $\Pi^p_{2t}$.

\begin{proposition}\label{prop:robust-general}
For all $t \geq 1$, \problem{Robust $t$-Reachability} is in $\Pi^p_{2 t}$.
\end{proposition}

\begin{proof}
One round of robustness can be viewed as follows: (a) Universally, the adversary chooses a sequence. (b) Existentially, the other player selects functions. (c) The system executes the functions according to the sequence.
Thus, it is in $\Pi^p_{2}$.
With $t$ rounds, we have $t$ of the three-step action and so the reachability problem is in $\Pi^p_{2t}$.
\end{proof}

Surprisingly, there is a stricter upper bound.

\begin{theorem}\label{thm:robustness-one}
For all $t \geq 1$, \problem{Robust $t$-Reachability} is in $\Pi^p_{2(t-1)}$ if the functions are arbitrary bounded-fan-in functions, the unbounded-fan-in OR, o the unbounded-fan-in AND.
\end{theorem}

\begin{proof}
Let $\syst{F} = \{ f_{i, j} \mid i \in [n], j\in [k] \}$ be a DT-BFDS and let $f$ be one of the functions.
Suppose $\config{c}$ and $\config{d}$ be configurations of $\syst{F}$ and we are testing if $\syst{F}$ can robustly transition from $\config{c}$ to $\config{d}$.
In other words, we are asking for all permutations $\pi$, there exist a function index set with which $\syst{F}$ takes $\config{c}$ to $\config{d}$ in one step.

The negation of the condition is:
\begin{description}
\item[(A)]
there exists a permutation $\pi$ such that regardless of permissible function choices, $\syst{F}$ fails to drive $\config{c}$ to $\config{d}$ in one step.
\end{description}
Let select the first position in which the production of $\config{d}$ fails.
Then we have $\syst{F}$ does not robustly produce $\config{d}$ from $\config{c}$ if and only if:
\begin{description}
\item[(B)]
there exists a permutation $\pi$ and a variable $z$ such that even if the production of $\config{d}$ is successful for all variables preceding $z$, the production of $\config{d}$ for $z$ fails.
\end{description}
By swapping $\pi$ and $z$, we get that the robust transformation fails if and only if:
\begin{description}
\item[(C)]
there exists a variable $z$ and a permutation $\pi$ such that if the production of $\config{d}$ is successful for all variables preceding $z$, the production of $\config{d}$ for $z$ fails.
\end{description}
We examine the functions for $z$ to find if there is a permutation $\pi$ to satisfy the condition.
If the examination is possible in polynomial-time, then we execute the test for all variables $z$.
If there is any one such $z$, we find that $\syst{F}$ fails the robustness test.

Now, let us turn to one $z$.
Let $\Phi$ be the $k$ functions for $z$.
Let $f$ be one function in $\Phi$.
Let $\alpha$ be $z$'s value in $\config{c}$ and let $\beta$ be $z$'s value in $\config{d}$.
Let $f'(x_1, \ldots, x_m)$ be $f$' projection of $f$ keeping only those on which $f$ is dependent.
Let $c_1, \ldots, c_m$ be the elements on $\config{c}$ corresponding to $x_1, \ldots, x_m$ and let $d_1, \ldots, d_m$ be the elements on $\config{d}$ corresponding to $x_1, \ldots, x_m$.
For each $i \in [m]$, if $c_i = d_i$ and $x_i \neq z$, fix the value of $x_i$ to $c_i$ in $f'$ and remove $x_i$ from $f'$.
Let $f''(y_1, \ldots, y_r)$ be the resulting function and let $a_1, \ldots, a_r$ and $b_1, \ldots, b_r$ be those corresponding to $y_1, \ldots, y_r$ in $c$'s and $d$'s, respectively.
If $f''$ is the constant function that produces the target value $\beta$, regardless of the permutation, $f$ is able to produce the target value for $z$.
By selecting $f$ among the functions in $\Phi$, the system can regardless of the permutation order, $\syst{F}$ is able to produce $\beta$.
Thus, $z$ is not a variable satisfying (C), and so we move to another $z$.

Also, $f''$ is the constant function that produces the opposite of the target value, regardless of the permutation, $f$ produces the non-target value, and so we can safely remove $f$ from consideration and ask if there is a permutation $\pi$ for which regardless of the choice of functions from the remaining functions ini $\Phi$, $\syst{F}$ fails.

Let $U$ be the set of all variables among $y_1, \ldots, y_r$ not equal to $z$.
In the case where $f$ is bounded-fan-in, define $W_f$ be the set of all pairs $(S, U-S)$ such that flipping the states of $S$ and then computing $f''$ produces the output not equal to $\beta$.
Then we have
\begin{description}
\item[(*)]
With respect to function $f$, a permutation $\pi$ fails to produce the correct value for $z$ or an earlier variable if for some $(S, T) \in W_f$, $\pi$ processes $S$ before $z$ and $T$ after $z$.
\end{description}
In the case where $f$ is the unbounded fan-in OR and $\beta = 1$, define $W_f = \{ (S, T) \}$ such that $S$ consists of the variables that turn from $1$ to $0$.
The property (*) holds for $W_f$.
In the case where $f$ is the unbounded fan-in OR and $\beta = 0$, let $W'_f = \{ (S, T) \}$ such that $S$ consists of the variables that turn from $1$ to $0$.
We have:
\begin{description}
\item[(**)]
With respect to function $f$, a permutation $\pi$ fails to produce the correct value for $z$ or an earlier variable if either some variable in $S$ goes after $z$ or some variable in $T$ goes before $z$.
\end{description}
From this observation, we construct $W_f = \{ (\emptyset, \{ s \} ) \mid s \in S \} \cup \{ (\{ t \}, \emptyset) \mid t \in T \}$.
Then, $W_f$ satisfies (*).

In the case where $f$ is the unbounded fan-in AND and $\beta = 0$, let $W_f = \{ (S, T) \}$ such that $S$ consists of the variables that turn from $1$ to $0$.
In the case where $f$ is the unbounded fan-in AND and $\beta = 1$, let $W'_f = \{ (S, T) \}$ such that $S$ consists of the variables that turn from $1$ to $0$ and $W_f = \{ (\emptyset, \{ s \} ) \mid s \in S \} \cup \{ (\{ t \}, \emptyset) \mid t \in T \}$.
The set $W_f$ and $W'_f$ respectively satisfy (*) and (**).

Let $f$ and $g$ be two distinct functions in $\Phi_0$.
From each $(S, T)$ in $W_f$ and each $(S', T')$ in $W_g$ we construct. pair $(S\cup S', T\cup T')$.
Then a permutation that processes $S\cup S'$ before $z$ and $T \cup T'$ after $z$ fails to produce the correct value for $z$ or some earlier variable if either $f$ or $g$ is the choice of function.
If $S\cup S'$ and $T \cup T'$ are disjoint, such a permutation exists.
So, we take one pair from $W_f$ for each $f \in \Phi$ and create a new pair, where the first element of the pair is the union of the first parts and the second element of the pair is the union of the second parts.
We collect only those with non-intersection first and second parts from such union pairs, and construct the collection $W_z$.
Then we have that
\begin{description}
\item[(***)]
for each pair $(S, T) \in W_z$, a permutation that executes $S$ before $z$ and $T$ after $z$ fails to produce the correct value for $z$ or some earlier variable regardless of which function $\Phi$ may act.
\end{description}
Thus, $\syst{F}$ does not robustly transform $\config{c}$ to $\config{d}$ if and only if $W_z$ is not empty for some $z$.

Let $q$ be the bound on the fan-in in the case where the functions are bounded-fan-in.
Then each $W_f$ has at most $2^q$ elements.
In the case where the functions are unbounded-fan-in ORs or ANDs, the cardinality of $W_f$ is at most $n$.
Thus, the cardinality of $W_z$ is at most $n(\max\{ 2^q, n \})^k$, which is $O(n^{k+1})$.
This implies that we can test the robustness in time polynomial in $n$.
This proves the theorem.
\end{proof}

The previous theorem shows that \problem{Robust $t$-Reachability} for some types of functions, including the unary functions, belongs to $\Pi^p_{2(t-1)}$.
What can we say about the problem of testing if a permutation exists that enables the system to reach the target configuration?
Answering the question is difficult.
We only show that the question concerning  $1$-choice unary systems is polynomial-time solvable.

\begin{definition}\label{def:permutation-existence}
Let $t\geq 1$.
\problem{$t$-Permutation Existence} is the following decision problem.
The input to the problem consists of an $(n, k)$-system $\syst{F}$, $n, k\geq 1$, and two configurations $\config{c}$ and $\config{d}$, such that $\syst{F}$ uses the arbitrary permutation scheme and updates either independently or in coordination.
The question is whether there exists a series of $t$ permutations $[\pi_1, \ldots, \pi_t]$ such that when $\syst{F}$ uses $\pi_1, \ldots, \pi_t$ in time steps $1, \ldots, t$, respectively, there exists a series of function choices that takes $\config{c}$ to $\config{d}$.
\end{definition}

\begin{theorem}\label{thm:1-choice-existence}
\problem{$1$-Permutation Existence} for $1$-choice systems is polynomial-time solvable if the functions are unary.
\end{theorem}

\begin{proof}
Let $\syst{F} = \{ f_1. \ldots, f_n \}$ be an $(n, 1)$ DT-BFDS for some $n\geq 1$.
Let $v_1, \ldots, v_n$ be $\syst{F}$'s nodes.
Suppose we want to ask if $\syst{F}$ can take $\config{c} = (c_1, \ldots, c_n)$ to $\config{d} = (d_1, \ldots, d_n)$ in one time step.
We introduce two predicates.
Let $\pi$ be a variable representing a permutation.
One predicate is $S(\pi, i), i \in [n]$, representing ``if $\pi$ allows $v_i$ to update its state to $d_i$.''
The other predicate is $A(\pi, i, j), i, j \in [n]$, representing ``$i$ appears before $j$ in $\pi$.''
The question at hand is if there is $\pi$ such that
\[
\Phi(\pi) \equiv S(\pi, 1) \wedge S(\pi, 2) \wedge \cdots \wedge S(\pi, n).
\]
Let $i \in [n]$ be an arbitrary index and let $f_i$ be the function for $x_i$.
Let $x_j$ be the input to the function $x_j, j \in [n]$.
We can make the following analysis.
\begin{itemize}
\item
Suppose $f_i$ is a positive unary function; that is, $f_i = x_j$.
\begin{itemize}
\item
If $i = j$ and $c_i = \overline{d_i}$, while $v_i$ must flip its state, the function $f_i$ retains the value.
This means that $\syst{F}$ cannot achieve the transition from $c_i$ to $d_i$ for $v_i$ regardless of the permutation $\pi$.
Thus, $\Phi(\pi)$ is false for all $\pi$.

\item
If $i = j$ and $c_i = d_i$, the function of $f_i$ produces $d_i$ regardless of what $\pi$ may be.
Thus, $S(\pi, i) = \TRUE$ for all $\pi$.

\item
If $i \neq j$ and $c_j = d_j \neq d_i$, we have $f_i(c_j) = f_i(d_j) \neq d_i$, so $v_i$ cannot acquire the value $d_i$.
This means $S(\pi, i)$ is false for all permutations $\pi$, and so $\Phi(\pi)$ is false for all $\pi$.

\item
If $i \neq j$ and $c_j = d_j = d_i$, we have $f_i(c_j) = f_i(d_j) = d_i$, so $v_i$ can acquire the value $d_i$ regardless of the choice of $\pi$.
Thus, we have $S(\pi, i) = \TRUE$ for all $\pi$.

\item
If $i \neq j$ and $c_j \neq d_j = d_i$, we have $f_i(c_j) \neq f_i(d_j) = d_i$.
This means that $v_i$ can achieve its goal only if and only if $v_j$ goes before $v_i$ and $v_j$ is able to achieve its goal.
Thus, we have
\[
S(\pi, i) \equiv S(\pi, j) \wedge A(\pi, j, i).
\]

\item
If $i \neq j$ and $d_i = c_j \neq d_j$, we have $f_i(c_j) = d_i \neq f_i(d_j)$.
This means that $v_i$ can achieve$v_i$ can acquire the value $d_i$ if and only if $v_i$ appears somewhere before $v_j$ in $\pi$.
Thus, we have
\[
S(\pi, i) \equiv A(\pi, i, j).
\]
\end{itemize}

\item
Suppose $f_i = \overline{x_j}$.
\begin{itemize}
\item
If $i = j$ and $c_i = d_i$, $f_i(c_i) = \overline{c_i} \neq d_i$.
This means that $S(\pi, i)$, and therefore, $\Phi(\pi)$ is false regardless of $\pi$.

\item
If $i = j$ and $c_i \neq d_i$, $v_i$ can achieve its desired value, and so $S(\pi, i) = \TRUE$ regardless of $\pi$.

\item
If $i \neq j$ and $c_j = d_j \neq d_i$, we have $f_i(c_j) = f_i(d_j) = d_i$.
Thus,$S(\pi, i) = \TRUE$ regardless of $\pi$.

\item
If $i \neq j$ and $c_j = d_j = d_i$, we have $f_i(c_j) = f_i(d_j) \neq d_i$, and so $S(\pi, i)$ is false for all $\pi$.
Thus, $\Phi(\pi)$ is false for all $\pi$.

\item
If $i \neq j$ and $c_j = d_i \neq d_j$, $d_i = f_i(d_j) \neq f_i(c_j)$.
This means that $v_i$ can achieve the target if and only if $S(\pi, j) = \TRUE$ and $v_j$ appears before $v_i$ in $\pi$.
Thus, we have
\[
S(\pi, i) \equiv S(\pi, j) \wedge A(\pi, j, i).
\]

\item
If $i \neq j$ and $c_j \neq d_j = d_i$, $f_i(c_j) = d_i \neq f_i(d_j)$.
This means that $v_i$ can achieve the target if and only if $v_i$ appears before $v_j$ in $\pi$.
Thus, we have
\[
S(\pi, i) \equiv A(\pi, i, j).
\]
\end{itemize}
\end{itemize}
Suppose the analysis for none of the $i \in [n]$ produces $\Phi(\pi) = $ false, in which case we have that the answer to the reachability question is false.

Otherwise, we replace for each $i \in [n]$, $S(\pi, i)$ with the formula we have obtained.
Since the formula $\Phi(\pi)$ is the conjunction, we can remove all $S(\pi, i)$ that is equal to $\TRUE$.
This reduces $\Phi(\pi)$ to the conjunction of at most $n$ terns of $A(\pi, i, j), i \neq j$.
In other words, $\Phi(\pi) = \TRUE$ if and only if $\pi$ satisfies all the ordering conditions $A(\pi, i, j)$ appearing in the formula.

Suppose there is an index $i$ that appears in the formula such that $i$ appears only in the form $A(\pi, i, j)$ for some $j \neq i$.
Then we can choose $\pi$ so that $i$ is the first among all the indices appearing the formula, and remove all such terms.
Suppose there is an index $j$ that appears in the formula such that $j$ appears only in the form $A(\pi, i, j)$ for some $i \neq j$.
Then we can choose $\pi$ so that $j$ is the last among all the indices appearing the formula, and remove all such terms.
We repeat the removal until there is no such $i$ or $j$.
Let $\Psi(\pi)$ be the resulting formula and let $K$ be the set of all indices appearing in the formula.
Because each index contributes at most one term $A(\pi, i, j)$ to $\Psi(\pi)$ and each index appears in two different ways, we have that
\begin{itemize}
\item
There are $K$ terns in the formula.
\item
Each index $i \in K$ appears exactly once as the middle term of $A(\pi, i, j)$ for all $j$.
\item
Each index $j \in K$ appears exactly once as the last term of $A(\pi, i, j)$ for all $i$.
\end{itemize}
These properties are representable as a $K$-node directed graph, where each node has one incoming edge and one outgoing edge.
This means that the graph has a directed cycle.
Let $[i_1, \ldots, i_m, i_1]$ be one such cycle.
Then to satisfy the conditions, we have that in $\pi$, 
$i_1$ must appear before $i_2$, $i_2$ before $i_3$, and so on, and $i_m$ before $i_1$.
This means that there is no permutation satisfying all the $m$ conditions.
Thus, there is a permutation satisfying all the conditions if and only if $\Phi(\pi)$ is empty. 

It is not hard to see that constructing $\Psi$ is constructible from $\syst{F}$ in time polynomial in $n$.
This proves the theorem.
\end{proof}

\begin{corollary}\label{coro:1-choice-existence}
For an arbitrary $k\geq 1$, \problem{$1$-Permutation Existence} for $k$-choice systems is polynomial-time solvable if the functions are unary and the function selections are in coordination.
\end{corollary}

\begin{proof}
We can think of an $(n, k)$ system that makes coordinate function selection as a group of $k$ $(n, 1)$-systems.
We examine the question for each $(n, 1)$ system.
\end{proof}

We question if Theorem~\ref{thm:1-choice-existence} extends to systems that make individual selections.

We are unsure of its answer.
Here is our present investigation.

Let $\syst{F} = \{ f_{i, j} \mid i \in [n], j \in [k] \}$ be an $(n, k)$ DT BFDS for some $n, k\geq 1$ whose update functions are positive unary.
Let $v_1, \ldots, v_n$ be the system's nodes.
Let $x_1, \ldots, x_n$ represent their states.
Suppose we want to test if $\syst{F}$ can take $\config{c} = (c_1, \ldots, c_n)$ to $\config{d} = (d_1, \ldots, d_n)$ in one time step with some update sequence $\pi$ and function selection.
For a permutation $\pi$, let $\Phi(\pi)$ represent the property
\begin{center}
given $\pi$ as the update sequence, $\syst{F}$ can select functions to produce $\config{d}$ from $\config{c}$ in one step. 
\end{center}
We decompose the predicate $\Phi(\pi)$ as the conjunction of node-wise predicates
\[
\Phi(\pi) = S(\pi, 1) \wedge \cdots \wedge S(\pi, n)
\]
where $S(\pi, i)$ means
\begin{center}
given $\pi$ as the update sequence, $\syst{F}$ can select the function for $v_i$ to produce $d_i$.
\end{center}
We observe the following:
\begin{itemize}
\item
For all $i \in [n]$, if one of the functions for $v_i$ takes $x_\ell$ as the input (and thus, outputs $x_\ell$) and $d_i = c_\ell = d_\ell$, for all $\pi$, $S(\pi, i) = \TRUE$ so long as $S(\pi, \ell) = \TRUE$, and so we can safely remove $S(\pi, i)$ from $\Phi(\pi)$.
\item
For all $i \in [n]$, if one of the functions for $v_i$ takes $x_i$ as the input (and thus, outputs $x_i$) and $c_i \neq d_i$, that function is useless for making the change from $c_i$ to $d_i$, and so we can remove it from consideration.
\end{itemize}
After these removals, if for some $i$, there is no function remaining, we assert that $\Phi(\pi) = \FALSE$ for all $\pi$, and so the answer to the existential question is negative.

Let us assume we have at least one function remaining for each remaining $i$.
Let $i$ be a remaining $i$.
Suppose that one remaining function for $i$ produces the value of some $x_\ell$.
We have $i \neq \ell$ and $c_\ell \neq d_\ell$.
If $c_i = d_i$, then $i$ cannot appear as $\ell$ for any other values of $i$ and so we can select the position of $i$ to be the first among all the remaining indices if $d_i = c_\ell$ and the last among all if $d_i = d_\ell$.
Thus, if $c_i = d_i$, we can remove it from consideration as $S(\pi, i) = \TRUE$.
We can execute the removal for such $i$ in an arbitrary order.

Let $I$ be the set of all remaining $i$.
Let $L(i)$ be the set of all $\ell$ such that the function that produces $x_\ell$ is one of the remaining functions for $I$.
We have for all $i \in I$,
\[
\emptyset \neq L(i) \subseteq I - \{ i \}
\]
and $c_i \neq d_i$.
We partition $I$ into two sets, $L_1$ and $L_2$, where $L_1$ is the set of all $i \in I$ such that $c_i = \TRUE$ (and so $d_i = \FALSE$) and $L_2$ is the set of all $i \in I$ such that $c_i = \FALSE$ (and so $d_i = \TRUE$).
We construct an edge-labeled multi-edge digraph $G = (I, E)$ such that there is an edge $(i, \ell)$ with label $i$ if $\ell \in J(i)$ and $d_i = c_\ell$ (and so $c_i = d_\ell$) and there is an edge $(\ell, i)$ with label $i$ if $\ell \in J(i)$ and $d_i = d_\ell$ (and so $c_i = c_\ell$).

Recall that we are viewing the problem of selecting a permutation $\pi$ that achieves the goal as the problem of selecting a permutation $\sigma$ over $I$ that achieves the goal for the nodes with indices in $I$.
We claim that the latter problem is equivalent to the problem of selecting a set $R$ of $\| I\|$ edges in $E$ such that
\begin{itemize}
\item
for each $i$, there is an edge $(p, q) \in R$ with $i$ as the label (that is, either $p = i$ or $q = i$), and
\item
the edge-induced subgraph of $G$ concerning $R$, $G|_R$, is cycle-free.
\end{itemize}
The reason that the claim holds is as follows.
Suppose there is a permutation $\sigma$ that achieves the goal.
There must be an accompanying function selection for $\sigma$.
For each $i \in I$, fix one such function selection and let $\lambda(i)$ to be such that $x_{\lambda(i)}$ is the variable the function uses.
Think of an edge between $i$ and $\lambda(i)$ with $i$ as the label.
where the edge's direction is from $i$ to $\lambda(i)$ if $\sigma(i)$ appears before $\lambda(i)$ in $\sigma$ and the direction is opposite otherwise.
Based on how we constructed $G$, the edge must belong to $G$.
Let $R$ be the edges we have thus chosen.
These edges clearly a part of $E$.
Since the direction of the edges respect $\sigma$, they induce no cycles.

On the other hand, suppose there is a selection $R$ that satisfies the two conditions.
Let $H = G|_R$ be the graph $R$ induces on $G$.
Since the selection induces no cycles, we can stratify the set $I$ according to the selection.
Level $0$ consists of all the nodes without incoming edges.
After collecting level-$0$ nodes, at each level $l \geq 1$, we collect all the nodes whose distance (i.e., the length of the longest path from any level-$0$ node) is equal to $l$.
We keep the process of adding levels until we have collected all the nodes in $I$.
Because the $H$ has no cycles, we can complete the process.
Think of $\sigma$ as a permutation that enumerates the nodes according to the stratification, where the level $0$ nodes appear in some order, the level $1$ nodes appear next in some order, etc.
For each node $i$, we have either an outgoing edge with $i$ as the label or an incoming edge with $i$ as the label, but no both.
For the former, let $\ell$ be the destination of the edge, and we make a function selection such that $x_\ell$ is the state the function uses, and for the latter, let $\ell$ be the origin of the edge , and we make a function selection such that $x_\ell$ is the state the function uses.
Then, the function selection is possible with respect to the permutation $\sigma$.

The latter interpretative problem is equivalent to:
\begin{itemize}
\item
selecting exactly one edge with $i$ as the label for each $i$ so that the chosen edges do not induce a cycle.
\end{itemize}

\paragraph{Arthur--Merlin}
Another possible twist to the robustness study is to use will use randomness for one player's choice, like in the Arthur--Merlin games and Merlin--Arthur games~\cite{bab:c:trading,gol-sip:c:private-coins}.
In the AM and MA models, Merlin acts as an omnipotent adversary, and Arthur counters with random choices to achieve the goal with high probability.

\begin{definition}\label{def:merlin-arthur}
\deftitle{Merlin-Arthur BFDS}
An $(n, k)$ Merlin-Arthur DT-BFDS $\syst{F} = \{ f_{i, j} \mid i\in[n], j\in[k] \}$ is an $(n, k)$ Merlin-Arthur DT-BFDS that operates as follows.
At each time step:
\begin{itemize}
\item
Merlin chooses an update sequence $\pi$; i.e., a permutation of $[n]$; then
\item
Arthur probabilistically chooses $(j_1, \ldots, j_n) \in [k]^n$ and then applies $\tilde{f}_{\pi(i),j_{\pi(i)}}$ for $i = 1, \ldots, n$.
\end{itemize}
We can also define an Arthur-Merlin version where the order of action between the two is in the reverse order.
Also, we can assign the selection of functions to Merlin and the sequence selection to Arthur.
\end{definition}

\section{Conclusion and Open Questions}\label{sec:open}

In this paper, we introduced the notion of DT-BFDS with uncertainty and proved some initial results.
There are many interesting open questions.
We hope to explore further these uncertainty models.
We have several open questions.

\begin{enumerate}
\item
\label{q:general-upper-bound}
\myemph{(Lowering Upper Bounds)~}
Proposition~\ref{prop:general-upper-bound} shows upper bounds for various structural problems.
Under what conditions can we lower the new upper bounds?

\item
\label{q:individual-time-step-restriction}
\myemph{(Equivalence Between Parallel and Permutation-list)~}
Corollary~\ref{coro:subsumption} shows relations among updating schedules, can the relations be equivalences or proper inclusions.

\item
\label{q:k+1-versus-k}
\myemph{($k+1$ Choices Versus $k$ Choices)~}
Theorem~ \ref{thm:coordinated-k-to-3} shows that in the case of coordinated updates, each $k$-choice system is simulate-able with a $3$-choice system.
Can we further reduce it to $2$-choice systems?
Also, can we show a similar result for other function selection schemes?

\item
\label{q:path-counting}
\myemph{(Path Counting)~}
Can we show that \problem{Path Counting} for \problem{$t$-Reachability} is $\sharpp$-complete for some $t$, perhaps by finding a witness-preserving reduction?

\item
\label{q:limited-nondeterminism-one}
\myemph{(Further Fewer Nodes with Multiple Choices)~}
Theorem~\ref{thm:four-nodes-unary} shows that only two nodes with multiple choices are necessary for connecting configuration nodes. 
Can the number be smaller than $2$?

\item
\label{q:graph-iso}
\myemph{(Characterizing Graph Isomorphism with or without Basis)~}
Can the Graph Isomorphism result in Corollary~\ref{coro:graph-iso} be an equivalence?

\item
\label{q:cyclic-BFDS-general}
\myemph{(Complexity Class Characterization of Cyclic DT-BFDS)~}
Is \problem{Reachability} for pure-cyclic DT-BFDS's that update in parallel and select functions in coordination complete for some complexity classes?

\item
\label{q:coordinated-npc-improve}
Theorem~\ref{thm:coordinated-npc} shows that $4$ choices are sufficient for\problem{$t$-Reachability} for parallel, coordinated models to be $\np$-complete.
Can we reduce the number from $4$ to a smaller number?

\item
\label{q:cyclic-length}
\myemph{(Minimum Cycle Length Calculation)~}
How complex is the problem of computing the cycle length for cyclic systems?

\item
\label{q:nondet-seq-complete}
\myemph{(Permutation-list/Arbitrary-permutation and $\pspace$-completeness)~} 
Is \problem{Reachability} for $1$-choice DT-BFDS that use either a list of permutations or an arbitrary permutation $\pspace$-complete with some choice of update functions?
Is \problem{$t$-Reachability} with the same setting $\np$-complete with some choice of update functions?

\item
\label{q:parallel-versus-sequential}
What is the relationship between the parallel, unary, multiple-choice DT-BFDS with the individual function selection and the sequential, unary, $1$-choice DT-BFDS with the permutation-list schedule or the arbitrary-permutation schedule?

\item
\label{q:all-sequences}
\myemph{(Checking If All Permutations Are Successful)~}
For \problem{$1$-Reachability}, the question of whether all sequences achieve the goal is in $\conp$.
Is \problem{$1$-Reachability} $\conp$-complete for models that use a permutation-list or an arbitrary permutation?

\item
\label{q:counting-sequences}
\myemph{($\sharpp$-completeness of Counting Permutations)~}
with the permutation-list or the arbitrary-permutation schedule, the problem of counting updating sequences that take the initial configuration to the target configuration is in $\sharpp$.
Can the problem be $\sharpp$-complete?

\item
\label{q:robustness}
\myemph{(Completeness of Robust $t$-Reachability)~}
Is \problem{Robust $t$-Reachability} $\Pi_{2(t-1)}^p$-complete if the functions are bounded-fan-in, the unbounded-fan-in ORs, or the unbounded-fan-in ANDs?
If we are to swap the quantifiers, will the corresponding problem become $\Sigma_{2(t-1)}^p$-complete?

\item
\label{q:2-choice-existence}
Theorem~\ref{thm:1-choice-existence} shows that the permutation existence problem for \problem{$1$-Reachability} regarding $1$-choice systems is polynomial-time solvable if the functions are unary. What is the complexity of the problem for $2$-choice systems?

\item
\label{q:ma-and-am}
\myemph{(Computable Problems with an AM or MA Framework)~}
What can we say about the computation power of Merlin-Arthur DT-BFDS and Arthur-Merlin DT-BFDS?
For example, does an AM protocol exist for \problem{Graph Isomorphism}?

\end{enumerate}


\end{document}